\documentclass[11pt]{amsart}

\pdfoutput=1

\usepackage{amssymb}
\usepackage{epsfig}
\usepackage{comment}
\usepackage{amsmath}
\usepackage{subcaption}
\usepackage[section]{placeins}
\usepackage{graphicx}
\usepackage{units}

\usepackage{amsmath,amsfonts, amssymb, graphicx, caption, refstyle, float}
\usepackage{algorithm,algorithmic}
\usepackage{varwidth}
\usepackage{color}
\usepackage{subcaption}
\usepackage{float}

     \def\bB{{\mbs{B}}}

    \def\bP{{\mbs{P}}}

      \def\fb{{\mbs{f}}}

    \def\bp{{\mbs{p}}} \def\bq{{\mbs{q}}} 
      \def\bu{{\mbs{u}}}
      
    \def\by{{\mbs{y}}} \def\bz{{\mbs{z}}}

\theoremstyle{definition}

\theoremstyle{remark}

    \newcommand{\mbs}[1]{\boldsymbol{#1}}

\begin{document}

\title{Model-Free Data-Driven Inelasticity}

\author{\tiny R.~Eggersmann$^1$, T.~Kirchdoerfer$^3$, S.~Reese$^1$, L.~Stainier$^2$ and M.~Ortiz$^{3,4}$}

\address
{
    $^1$Institute of Applied Mechanics,
    RWTH Aachen University,
    Mies-van-der-Rohe-Str.~1, D-52074 Aachen, Germany.
    \newline\indent
    $^2$Institute of Civil and Mechanical Engineering (UMR 6183 CNRS/ECN/UN),
    Ecole Centrale Nantes, 1 rue de la Noe, F-44321 Nantes, France.
    \newline\indent
    $^3$Division of Engineering and Applied Science,
    California Institute of Technology,
    1200 E.~California Blvd., Pasadena, CA 91125, USA.
    \newline\indent
    $^4$ Hausdorff Center for Mathematics, Universit\"at Bonn,
    Endenicher Allee 60, 53115 Bonn, Germany.
}

\email{ortiz@caltech.edu}

\begin{abstract}
We extend the Data-Driven formulation of problems in elasticity of Kirchdoerfer and Ortiz \cite{Kirchdoerfer:2016} to inelasticity. This extension differs fundamentally from Data-Driven problems in elasticity in that the material data set evolves in time as a consequence of the history dependence of the material. We investigate three representational paradigms for the evolving material data sets: i) {\sl materials with memory}, i.~e., conditioning the material data set to the past history of deformation; ii) {\sl differential materials}, i.~e., conditioning the material data set to short histories of stress and strain; and iii) {\sl history variables}, i.~e., conditioning the material data set to {\sl ad hoc} variables encoding partial information about the history of stress and strain. We also consider combinations of the three paradigms thereof and investigate their ability to represent the evolving data sets of different classes of inelastic materials, including viscoelasticity, viscoplasticity and plasticity. We present selected numerical examples that demonstrate the range and scope of Data-Driven inelasticity and the numerical performance of implementations thereof.
\end{abstract}

\maketitle

\section{Introduction}

Kirchdoerfer and Ortiz \cite{Kirchdoerfer:2016, Kirchdoerfer:2017, Kirchdoerfer:2018} and Conti {\sl et al.} \cite{Conti:2018} have recently proposed a new class of problems in static and dynamic elasticity, referred to as {\sl Data-Driven problems}, defined on the space of strain-stress field pairs, or phase space. The problems consist of minimizing the distance between a given material data set and the subspace of compatible strain fields and stress fields in equilibrium. They find that the classical solutions are recovered in the case of linear elasticity and identify conditions for convergence of Data-Driven solutions corresponding to sequences of material data sets. Data-Driven elasticity effectively reformulates the classical initial-boundary-value problem of elasticity directly from material data, thus bypassing the empirical material modelling step altogether. By eschewing empirical models, material modelling empiricism, modelling error and uncertainty are eliminated entirely and no loss of experimental information is incurred.

It should be noted that the use of material data as a basis for constitutive modeling is classical and remains the subject of extensive ongoing research. There is a vast body of literature devoted to that subject, including recent developments based on statistical learning, model and data reduction, nonlinear regression, and others, which would be too lengthy to enumerate here. It bears emphasis, that what sets the present approach apart from these other approaches is that we reformulate the classical boundary value problems of mechanics, including inelasticity and approximations thereof, directly on the basis of the material data, without any attempt at modeling the data or performing any form of data reduction or manipulation.

A natural extension of the Data-Driven paradigm concerns inelastic materials whose response is irreversible and history dependent. The theory of {\sl materials with memory} furnishes the most general representation of such materials. According to Rivlin \cite{Rivlin:1972}:
\begin{quotation}
{\sl "The characteristic property of inelastic solids which distinguishes them from elastic solids is the fact that the stress measured at time t depends not only on the instantaneous value of the deformation but also on the entire history of deformation."}
\end{quotation}
The origins of the theory may be traced to a series of papers by Green and Rivlin starting in 1957 \cite{Green:1957, Green:1959, Green:1960}, who proposed the use of hereditary constitutive laws, originally developed by Boltzman \cite{Boltzmann:1874} and Volterra \cite{Volterra:1909} in the linear case, for the description of non-linear viscoelastic materials as an alternative to models using constitutive equations of the rate type \cite{Rivlin:1955}. The hereditary functional approach to inelasticity was introduced into thermodynamics by Coleman \cite{Coleman:1964a}. A linearization of Green and Rivlin's theory was developed by Pipkin and Rivlin \cite{Pipkin:1961}. Rheological properties of solids often have a {\sl fading memory property}, enunciated by Truesdell \cite{Truesdell:1965} as:
\begin{quotation}
{\sl "Events which occurred in the distant past have less influence in determining the present response than those which occurred in the recent past"}.
\end{quotation}
The concept of fading memory was formalized by Coleman and Noll \cite{Coleman:1961a, Coleman:1961b} as a continuity property of the hereditary functional and subsequently extended by Wang \cite{Wang:1965a, Wang:1965b}, Perzyna \cite{Perzyna:1967} and others.

Other general representations of inelasticity are based on continuum thermodynamics with internal variables (cf., e.~g., \cite{Rice:1975}). These representations replace an explicit dependence on history by a dependence on the effects of history, i.~e., the current microstructure of the material element. The variables used to describe that microstructure are referred to as internal variables. Together with the state of stress or deformation and a thermodynamic variable such as temperature or entropy, they define the local state of a material element. Such models were introduced for viscoelastic deformation by Eckart \cite{Eckart:1948}, Meixner \cite{Meixner:1953}, Biot \cite{Biot:1954} and Ziegler \cite{Ziegler:1958}, and have been extensively studied since \cite{Coleman:1963, Schapery:1964, Valanis:1966, Coleman:1967, Lubliner:1972, Lubliner:1973}. The foundations underlying the memory-functional and the internal-variable formalisms were critically reviewed by Kestin and Rice \cite{Kestin:1970}. The correspondence and, in some cases, equivalence between the material-with-memory, internal variable and differential formulations of inelasticity have also been extensively investigated \cite{Coleman:1967, Valanis:1967, Lubliner:1969, Lubliner:1973}.

In the context of Data-Driven inelasticity, the representational paradigms just outlined translate into corresponding representational paradigms for the material data set. Specifically, we identify the material data set $D(t)$ at time $t$ with the collection of stress-strain pairs $(\epsilon(t),\sigma(t))$ that are attainable by the material at that time. For inelastic materials, $D(t)$ depends on the past history $\{(\epsilon(s),\sigma(s))\}_{s < t}$ of stress and strain. The central issue of Data-Driven inelasticity thus concerns the formulation of rigorous yet practical representational paradigms for the evolving material data set. The practicality of the representation revolves around the amount of data that needs to be carried, or generated, along with the calculations. By rigorous we specifically mean representations that result, albeit at increasing computational cost, in convergent approximations.

We specifically consider three representational paradigms: i) {\sl materials with memory}, i.~e., conditioning the material data set to the past history of deformation; ii) {\sl differential materials}, i.~e., conditioning the material data set to short histories of stress and strain; and iii) {\sl history variables}, i.~e., conditioning the material data set to {\sl ad hoc} variables encoding partial information about the history of stress and strain. We also consider combinations of the three paradigms thereof and investigate their ability to represent the evolving data sets of different classes of inelastic materials, including viscoelasticity, viscoplasticity and plasticity. The resulting Data-Driven inelasticity problems then consist of minimizing distance in phase space between the evolving data set and a time-dependent constraint set. We additionally concern ourselves with the numerical implementation and convergence characteristics of the resulting Data-Driven schemes.

We structure the paper as follows. In Section~\ref{L4cpYd} we succinctly summarize the Data-Driven approach to elasticity by way of background and in order to set  essential notation. Extensions to inelasticity predicated on various representations of the material data set are put forth and developed in Section~\ref{4qR2FP}. In Section~\ref{IXEV5B} we present selected examples of application to viscoelastic solids that demonstrate the suitability of differential representations of the material data set and the performance of the resulting Data-Driven schemes. Further examples of application are presented in Section~\ref{X0aZjS} that demonstrate how hybrid differential/history variable representations of the material data set can be used to account for hardening plasticity. Finally, an extended discussion of possible extensions and alternative approaches is presented in Section~\ref{y1HJx6}.

\section{Background: Data-Driven elasticity}
\label{L4cpYd}

We begin by recalling the Data-Driven reformulation of elasticity \cite{Kirchdoerfer:2016, Kirchdoerfer:2017} as a basis for subsequent generalizations to inelasticity. For simplicity, we consider discrete, or discretized, systems consisting of $N$ nodes and $M$ material points. The system undergoes displacements $\bu = \{\bu_a\}_{a=1}^N$, with $\bu_a \in \mathbb{R}^{n_a}$ and $n_a$ the dimension of the displacement at node $a$, under the action of applied forces $\fb = \{\fb_a\}_{a=1}^N$, with $\fb_a \in \mathbb{R}^{n_a}$. The internal state of the system is characterized by local stress and strain pairs $\{(\mbs{\epsilon}_e, \mbs{\sigma}_e)\}_{e=1}^M$, with $\mbs{\epsilon}_e, \mbs{\sigma}_e \in \mathbb{R}^{m_e}$ and $m_e$ the dimension of stress and strain at material point $e$. We regard $\bz_e = (\mbs{\epsilon}_e, \mbs{\sigma}_e)$ as a point in a local phase space $Z_e = \mathbb{R}^{m_e} \times \mathbb{R}^{m_e}$ and $\bz = \{(\mbs{\epsilon}_e, \mbs{\sigma}_e)\}_{e=1}^M$ as a point in the global phase space $Z = Z_1 \times \cdots \times Z_M$.

The internal state of the system is subject to the compatibility and equilibrium constraints of the general form
\begin{subequations}\label{XUI9Eu}
\begin{align}
    & \label{Bd2bYR}
    \mbs{\epsilon}_e = \bB_e \bu ,
    \quad e=1,\dots,M,
    \\ & \label{j8dCjM}
    \sum_{e=1}^M w_e \bB_e^T \mbs{\sigma}_e = \fb ,
\end{align}
\end{subequations}
where $\{w_e\}_{e=1}^M$ are elements of volume and $\bB_e$ is a discrete strain operator for material point $e$. We note that constraints (\ref{XUI9Eu}) are universal, or material-independent. They define a subspace, or constraint set,
\begin{equation}\label{DWdog7}
    E = \{ \bz \in Z \, : \, (\ref{Bd2bYR}) \text{ and } (\ref{j8dCjM}) \} ,
\end{equation}
consisting of all compatible and equilibrated internal states. In (\ref{DWdog7}) and subsequently, the symbol $:$ is used to mean 'given' or 'subject to' or 'conditioned to'. Within this subspace, the internal state satisfies the power identity
\begin{equation}
    \fb \cdot \bu
    =
    \sum_{e=1}^M w_e \, \mbs{\sigma}_e \cdot \mbs{\epsilon}_e .
\end{equation}

In classical elasticity, the problem (\ref{XUI9Eu}) is closed by appending local material laws, e.~g., functions of the general form
\begin{equation}\label{d6cdNV}
    \mbs{\sigma}_e = \hat{\mbs{\sigma}}_e(\mbs{\epsilon}_e) ,
    \quad
    e = 1,\dots, m ,
\end{equation}
where $\hat{\mbs{\sigma}}_e : \mathbb{R}^{m_e} \to \mathbb{R}^{m_e}$. However, often material behavior is only known through a material data set $D_e$ of points $\bz_e = (\mbs{\epsilon}_e,\mbs{\sigma}_e) \in Z_e$ obtained experimentally or by some other means. Again, the conventional response to this situation is to deduce a material law $\hat{\mbs{\sigma}}_e$ from the data set $D_e$ by some appropriate means, thus reverting to the classical setting (\ref{d6cdNV}).

The Data-Driven reformulation of the classical problems of mechanics consists of formulating boundary-value problems directly in terms of the material data, thus entirely bypassing the material modeling step altogether \cite{Kirchdoerfer:2016}. A class of Data-Driven problems consists of finding the compatible and equilibrated internal state $\bz \in E$ that minimizes the distance to the global material data set $D = D_1 \times \cdots \times D_M$. To this end, we metrize the local phase spaces $Z_e$ by means of norms of the form
\begin{equation}\label{fou9Oe}
    | \bz_e |_e
    =
    \left(
        \mathbb{C}_e \mbs{\epsilon}_e \cdot \mbs{\epsilon}_e
        +
        \mathbb{C}_e^{-1} \mbs{\sigma}_e \cdot \mbs{\sigma}_e
    \right)^{1/2} ,
\end{equation}
for some symmetric and positive-definite matrices $\{\mathbb{C}_e\}_{e=1}^M$, with corresponding distance
\begin{equation}
    d_e(\bz_e,\by_e) = | \bz_e - \by_e |_e ,
\end{equation}
for $\by_e, \bz_e \in Z_e$. The local norms induce a metrization of the global phase $Z$ by means of the global norm
\begin{equation}\label{9oakLa}
    | \bz |
    =
    \Big( \sum_{e=1}^m w_e | \bz_e |_e^2 \Big)^{1/2} ,
\end{equation}
with associated distance
\begin{equation}
    d(\bz,\by) = |\bz - \by|,
\end{equation}
for $\by, \bz \in Z$. The distance-minimizing Data-Driven problem is, then,
\begin{equation}\label{K5b6z0}
    \min_{\by \in D} \min_{\bz \in E} d(\bz,\by)
    =
    \min_{\bz \in E} \min_{\by \in D} d(\bz,\by) ,
\end{equation}
i.~e., we wish to find the point $\by \in D$ in the material data set that is closest to the constraint set $E$ of compatible and equilibrated internal states or, equivalently, we wish to find the compatible and equilibrated internal state $\bz \in E$ that is closest to the material data set $D$.

We emphasize that the local material data sets can be graphs, point sets, 'fat sets' and ranges, or any other arbitrary set in phase space. Evidently, the classical problem is recovered if the local material data sets are chosen as
\begin{equation}
    D_e
    =
    \{
        (\mbs{\epsilon}_e,\hat{\mbs{\sigma}}_e(\mbs{\epsilon}_e))
    \} ,
\end{equation}
i.~e., as {\sl graphs} in $Z_e$ defined by the material law (\ref{d6cdNV}). Thus, the Data-Driven reformulation (\ref{K5b6z0}) extends---and subsumes as special cases---the classical problems of mechanics.

We note that, for fixed $\by \in D$, the closest point projection $\bz = P_E \by$ onto $E$ follows by minimizing the quadratic function $d^2(\cdot,\by)$ subject to the constraints (\ref{XUI9Eu}). The compatibility constraint (\ref{Bd2bYR}) can be enforced directly by introducing a displacement field $\bu$. The equilibrium constraint (\ref{j8dCjM}) can then be enforced by means of Lagrange multipliers $\mbs{\lambda}$ representing virtual displacements of the system. With $\by \equiv \{(\mbs{\epsilon}'_e, \mbs{\sigma}'_e)\}_{e=1}^M$ given, e.~g., from a previous iteration, the corresponding Euler-Lagrange equations are \cite{Kirchdoerfer:2016}
\begin{subequations}
\begin{align}
    &
    \Big(\sum_{e=1}^M w_e \bB_e^T \mathbb{C}_e \bB_e \Big) \bu
    =
    \sum_{e=1}^M w_e \bB_e^T \mathbb{C}_e \mbs{\epsilon}'_e ,
    \\ &
    \Big(\sum_{e=1}^M w_e \bB_e^T \mathbb{C}_e \bB_e \Big) \mbs{\lambda}
    =
    \fb - \sum_{e=1}^M w_e \bB_e^T \mbs{\sigma}'_e ,
\end{align}
\end{subequations}
which define two standard linear displacement problems. The closest point $\bz = P_E \by \in E$ then follows as
\begin{subequations}
\begin{align}
    &
    \mbs{\epsilon}_e = \bB_e \bu ,
    \quad
    e = 1,\dots,M ,
    \\ &
    \mbs{\sigma}_e = \mbs{\sigma}'_e + \mathbb{C}_e \bB_e \mbs{\lambda} ,
    \quad
    e = 1,\dots,M .
\end{align}
\end{subequations}
A simple Data-Driven solver consists of the fixed point iteration \cite{Kirchdoerfer:2016}
\begin{equation}\label{K1TK7z}
    \bz_{j+1} = P_E P_D \bz_j ,
\end{equation}
for $j = 0,1,\dots$ and $\bz_0 \in Z$ arbitrary, where $P_D$ denotes the closest point projection in $Z$ onto $D$. Iteration (\ref{K1TK7z}) first finds the closest point $P_D \bz_j$ to $\bz_j$ on the material data set $D$ and then projects the result back to the constraint set $E$. The iteration is repeated until $P_D \bz_{j+1} = P_D \bz_j$, i.~e., until the data association to points in the material data set remains unchanged.

The convergence properties of the fixed-point solver (\ref{K1TK7z}) have been investigated in \cite{Kirchdoerfer:2016}. The Data-Driven paradigm has been extended to dynamics \cite{Kirchdoerfer:2018}, finite kinematics \cite{Nguyen:2018} and objective functions other than phase-space distance can be found in \cite{Kirchdoerfer:2017}. The well-posedness of Data-Driven problems and properties of convergence with respect to the data set have been investigated in \cite{Conti:2018}.

\section{Extension to inelasticity}
\label{4qR2FP}

A natural extension of the Data-Driven paradigm just described concerns inelastic materials whose response is irreversible and history dependent. The equilibrium boundary-value problem for these materials is, therefore, time dependent. For simplicity, we restrict attention to time-discrete formulations and seek to approximate solutions at times $t_0$, $t_1$, $\dots$, $t_k$, $t_{k+1}$, $\dots$. In this setting, the compatibility and equilibrium constraints (\ref{XUI9Eu}) become
\begin{subequations}\label{QB7839}
\begin{align}
    & \label{OJiD0T}
    \mbs{\epsilon}_{e,k+1} = \bB_e \bu_{k+1} ,
    \quad e=1,\dots,M,
    \\ & \label{R2Ku3B}
    \sum_{e=1}^M w_e \bB_e^T \mbs{\sigma}_{e,k+1} = \fb_{k+1} ,
\end{align}
\end{subequations}
where $\bu_{k+1}$, $\fb_{k+1}$, $\mbs{\epsilon}_{k+1}$ and $\mbs{\sigma}_{k+1}$ are the displacements, forces, strains and stresses at time $t_{k+1}$, respectively. The constraints (\ref{QB7839}) define the constraint set
\begin{equation}
    E_{k+1} = \{ \bz \in Z \, : \, (\ref{OJiD0T}) \text{ and } (\ref{R2Ku3B}) \} ,
\end{equation}
which is now time-dependent on account of the time-dependency of the applied loads.

In addition, the instantaneous response of inelastic materials is characterized by its dependence on the past history of deformation. By virtue of this history dependence, the set of stress-strain pairs attainable at a material point depends itself on time. We specifically define the instantaneous local material data set as
\begin{equation}\label{PS6guZ}
    D_{e,k+1}
    =
    \{
        (\mbs{\epsilon}_{e,k+1}, \mbs{\sigma}_{e,k+1}) \, : \,
        \text{past local history}
    \} ,
\end{equation}
i.~e., the set of local stress-strain pairs $(\mbs{\epsilon}_{e,k+1}, \mbs{\sigma}_{e,k+1})$ attainable at time $t_{k+1}$ at material point $e$ given the past history of the material point. We additionally define a global material data set at time $t_{k+1}$ as $D_{k+1} = D_{1,k+1}$ $\times$ $\cdots$ $\times$ $D_{M,k+1}$.

With these definitions, the Data-Driven problem of inelasticity is
\begin{equation}\label{yKEW6b}
    \min_{\by \in D_{k+1}} \min_{\bz \in E_{k+1}} d(\bz_{k+1},\by_{k+1})
    =
    \min_{\bz \in E_{k+1}} \min_{\by \in D_{k+1}} d(\bz_{k+1},\by_{k+1}) ,
\end{equation}
i.~e., we wish to find the point $\by_{k+1}$ in the material data set $D_{k+1}$ at time $t_{k+1}$ that is closest to the constraint set $E_{k+1}$ at time $t_{k+1}$ or, equivalently, we wish to find the internal state $\bz_{k+1}$ in the constraint set $E_{k+1}$ at time $t_{k+1}$ that is closest to the material data set $D_{k+1}$ at time $t_{k+1}$. Evidently, the inelastic Data-Driven problem (\ref{yKEW6b}) represents a natural extension of the elasticity Data-Driven problem (\ref{K5b6z0}) in which both the constraint set and the material data set are a function of time.

The central challenge now is to formulate rigorous yet practical means of characterizing the history dependence of the local material data sets $D_{e,k+1}$, eq.~(\ref{PS6guZ}). As noted in the introduction, inelastic material behavior can alternatively be described by means of hereditary laws, within the general framework of materials with memory, rheological and thermodynamical models based on internal variables, by means of so-called differential models and by other means. These constitutive formulations give rise to corresponding representational paradigms in the context of Data-Driven inelasticity, which we elucidate next.

\subsection{General materials with memory}

A general material with memory is a material whose state of stress is a function of the past history of strain, i.~e.,
\begin{equation}\label{9qwUSn}
    \mbs{\sigma}_e(t)
    =
    \hat{\mbs{\sigma}}_e(\{\mbs{\epsilon}_e(s)\}_{s \leq t}) ,
\end{equation}
where $\mbs{\sigma}_{e}(t)$ is the stress at material point $e$ and time $t$, $\{\mbs{\epsilon}_e(s)\}_{s \leq t}$ is the corresponding history of strain prior to $t$ and $\hat{\mbs{\sigma}}_e$ is a hereditary functional. For linear rheological materials, $\hat{\mbs{\sigma}}_e$ takes the form of a hereditary or Duhamel integral expressed in terms of a relaxation kernel \cite{Flugge:1975}.

In a discrete setting, (\ref{9qwUSn}) can be approximated as
\begin{equation}\label{mKGT32}
    \mbs{\sigma}_{e,k+1}
    =
    \hat{\mbs{\sigma}}_e(\{\mbs{\epsilon}_{e,l}\}_{l\leq k+1})
\end{equation}
where $\mbs{\sigma}_{e,k+1}$ is the stress at material point $e$ at time $t_{k+1}$, $\{\mbs{\epsilon}_{e,l}\}_{l\leq k+1}$ is the strain history of material point $e$ up to time $t_{k+1}$ and $\hat{\mbs{\sigma}}_e$ is a discrete hereditary function. In this representation, the local material data sets (\ref{PS6guZ}) take the form
\begin{equation}\label{2oKN63}
    D_{e,k+1}
    =
    \{
        (\mbs{\epsilon}_{e,k+1}, \mbs{\sigma}_{e,k+1}) \, : \,
        \{\mbs{\epsilon}_{e,l}\}_{l\leq k}
    \} ,
\end{equation}
i.~e., they consist of pairs $(\mbs{\epsilon}_{e,k+1}, \mbs{\sigma}_{e,k+1})$ of stress and strain known to be attainable at time $t_{k+1}$ given the past history $\{\mbs{\epsilon}_{e,l}\}_{l\leq k}$. In particular, we note that the material data set at time $t_{k+1}$ depends on the entire history of strain up to and including time $t_k$.

As noted in the introduction, materials often exhibit a fading memory property whereby their instantaneous behavior is a function primarily of the recent state history and is relatively insensitive to the distant past history. Examples include viscoelastic materials exhibiting relaxation and bounded creep. For those materials, the strain history in (\ref{mKGT32}) can be truncated beyond a certain decay time, which simplifies the parametrization of the local material data sets $D_{e,k+1}$. These simplifications notwithstanding, keeping track of long deformation histories, and sampling material behavior conditioned to them, may be challenging and onerous even for materials with fading memory.

\subsection{Internal variable formalism}

Thermodynamic models based on internal variables are often used to characterize inelasticity and history dependence. In these models, the state at a material point $e$ is described in terms of, e.~g., its strain, temperature and an additional array of auxiliary variables $\bq_e$ variables, or internal variables. Thermal processes are beyond the scope of this paper and we shall omit explicit reference to temperature and other thermodynamic variables for simplicity.

In order to describe the behavior of the material, we may assume a Helmholtz free energy $F_e(\mbs{\epsilon}_e, \bq_e)$, with corresponding equilibrium relations
\begin{subequations}\label{O4cXId}
\begin{align}
    & \label{e36oQ2}
    \mbs{\sigma}_e(t)
    =
    D_1F_e(\mbs{\epsilon}_e(t), \bq_e(t)) ,
    \\ & \label{e33BDo}
    \bp_e(t)
    =
    -
    D_2F_e(\mbs{\epsilon}_e(t), \bq_e(t)) ,
\end{align}
\end{subequations}
where $\bp_e$ are thermodynamic driving forces conjugate to $\bq_e$ and $D_1F_e$ and $D_2F_e$ denote the derivatives of $F_e$ with respect to strain and internal variables, respectively. In addition, the evolution of the internal variables is governed by kinetic relations of the form
\begin{equation}\label{Lg7W5y}
    D\psi_e(\dot{\bq}_e(t))
    +
    D_2F_e(\mbs{\epsilon}_e(t), \bq_e(t))
    =
    \mbs{0} ,
\end{equation}
where $\psi_e$ is a dissipation function and $D\psi_e$ its derivative.

In a time-discrete setting, the evolution of the internal variables is governed by incremental kinetic relations, e.~g., of the form \cite{Ortiz:1999}
\begin{equation}\label{ZnF8G1}
    D\psi_e(\frac{\bq_{e,k+1}-\bq_{e,k}}{t_{k+1}-t_k})
    +
    D_2F_e(\mbs{\epsilon}_{e,k+1}, \bq_{e,k+1})
    =
    \mbs{0} ,
\end{equation}
and the stress-strain relations (\ref{e36oQ2}) specialize to
\begin{equation}\label{0BDNIT}
    \mbs{\sigma}_{e,k+1}
    =
    D_1F_e(\mbs{\epsilon}_{e,k+1}, \bq_{e,k+1}) .
\end{equation}
Eqs.~(\ref{ZnF8G1}) and (\ref{0BDNIT}) define a close system of equations that can be solved for $\bq_{e,k+1}$ and $\mbs{\sigma}_{e,k+1}$ given $\mbs{\epsilon}_{e,k+1}$ and $\bq_{e,k}$. The corresponding material data set admits the representation
\begin{equation}\label{7cY9he}
    D_{e,k+1}
    =
    \{
        (\mbs{\epsilon}_{e,k+1}, \mbs{\sigma}_{e,k+1}) \, : \,
        \bq_{e,k}
    \} ,
\end{equation}
i.~e., $D_{e,k+1}$ is the set of all stress and strain pairs $(\mbs{\epsilon}_{e,k+1}, \mbs{\sigma}_{e,k+1})$ accessible to the material given the prior internal state $\bq_{e,k}$.

\subsection{Relation between the internal variable and hereditary representations}
\label{43VIZg}

The internal variable formalism, eqs.~(\ref{O4cXId}) and (\ref{Lg7W5y}), may be regarded as a convenient device for defining hereditary laws of the form (\ref{9qwUSn}). Thus, let \begin{equation}
    \bq_e(t)
    =
    \hat{\bq}_e(\{\mbs{\epsilon}_e(s)\}_{s \leq t})
\end{equation}
denote the solution of (\ref{Lg7W5y}), regarded as a system of ordinary differential equations in $\bq_e(t)$. Inserting into (\ref{e36oQ2}), we obtain the hereditary law
\begin{equation}
    \mbs{\sigma}_e(t)
    =
    D_1F_e(\mbs{\epsilon}_e(t), \hat{\bq}_e(\{\mbs{\epsilon}_e(s)\}_{s \leq t})) ,
\end{equation}
which, evidently, is a particular case of (\ref{9qwUSn}).

In the time-discrete setting, the internal variable formalism, eqs.~(\ref{ZnF8G1}) and (\ref{0BDNIT}), may also be regarded as a means of defining time-discrete hereditary laws of the form (\ref{mKGT32}). Thus, solving (\ref{ZnF8G1}) for the interval variables gives a relation
\begin{equation}\label{iQG1K4}
    \bq_{e,k+1}
    =
    \bP_e(\bq_{e,k}, \mbs{\epsilon}_{e,k+1}) ,
\end{equation}
where $\bP_e$ plays the role of a {\sl propagator}. Inserting into (\ref{0BDNIT}), we further obtain the stress-strain relation
\begin{equation}\label{fYFK2S}
    \mbs{\sigma}_{e,k+1}
    =
    D_1F_e(\mbs{\epsilon}_{e,k+1}, \bP_e(\bq_{e,k}, \mbs{\epsilon}_{e,k+1})) ,
\end{equation}
conditioned to the prior internal state $\bq_{e,k}$. Iterating this relation, we obtain
\begin{equation}
\begin{split}
    \mbs{\sigma}_{e,k+1}
    & =
    D_1F_e
    (
        \mbs{\epsilon}_{e,k+1},
        \bP_e(\bP_e(\bP_e(\cdots, \mbs{\epsilon}_{e,k-1}), \mbs{\epsilon}_{e,k}), \mbs{\epsilon}_{k+1})
    )
    \\ & \equiv
    \hat{\mbs{\sigma}}_e(\{\mbs{\epsilon}_{e,l}\}_{l\leq k+1}) ,
\end{split}
\end{equation}
which defines a discrete hereditary law of the form (\ref{mKGT32}) for the stresses as a function of the past history of strain. However, instead of the general history parametrization (\ref{2oKN63}) the material data set now admits the more explicit representation (\ref{7cY9he}), which greatly reduces the complexity of the parametrization of the material data set relative to that based on a general hereditary framework.

\subsection{History variables}

Despite its appeal, the essential conceptual drawback of the internal variable formalism is that the internal variable set is often not known or is the result of modeling assumptions. The efficiency of the internal variable parametrization can be retained, while eschewing {\sl ad hoc} modeling assumptions, simply by reinterpreting internal variables as history variables. Contrary to internal variables, history variables need not have a specific physical meaning and their function is simply to record partial information about the history of the material.

By way of motivation, we may iterate the update (\ref{iQG1K4}) to obtain the relation
\begin{equation}
    \bq_{e,k}
    =
    \bP_e(\bP_e(\bP_e(\cdots, \mbs{\epsilon}_{e,k-2}), \mbs{\epsilon}_{e,k-1}), \mbs{\epsilon}_{k})
    \equiv
    \hat{\bq}_e(\{\mbs{\epsilon}_{e,l}\}_{l\leq k}) ,
\end{equation}
which gives the internal variables at $t_{k}$ as a function of the strain history up to and including $t_k$. More generally, we may consider history variables of the form
\begin{equation}\label{RTL6EL}
    \bq_{e,k}
    =
    \hat{\bq}_e
    (
        \{\mbs{\epsilon}_{e,l}\}_{l\leq k} ,
        \{\mbs{\sigma}_{e,l}\}_{l\leq k}
    ) ,
\end{equation}
i.~e., functions of the stress and strain histories up to and including $t_k$. Implicit in the internal variable framework is that the current material data set $D_{e,k+1}$ depends on the deformation history only through a reduced set of history-dependent internal variables $\bq_{e,k}$, eq.~(\ref{7cY9he}).

The paradigm shift now consists of regarding the variables $\bq_{e,k}$ not as physical variables but as {\sl ad hoc} history variables that record and store partial information about the past internal history of the material point. Thus, the history variables $\bq_{e,k}$ at time $t_k$ are the result of applying {\sl ad hoc} history functionals $\hat{\bq}_e$ to the prior history of stress and strain. The history functionals query that history and extract and record selected information. The history information is then used to condition and parametrize the material data sets as in (\ref{7cY9he}). However, in the new reinterpretation (\ref{7cY9he}) represents the set of all known stress and strain pairs $(\mbs{\epsilon}_{e,k+1}, \mbs{\sigma}_{e,k+1})$ consistent with all past stress and strain histories for which the chosen history functionals $\hat{\bq}_e$ evaluate to $\bq_{e,k}$.

Importantly, the choice of history variables is no longer a matter of material modeling, as is the case for internal variables, but a question of approximation theory. Specifically, the aim is to produce sequences of history functionals that constrain arbitrary histories of stress and strain increasingly tightly, and exactly in the limit.
In particular, the sequence of Data-Driven solutions constrained by an increasing number of history variables should converge to the exact Data-Driven solution corresponding to (\ref{2oKN63}). In practice, the central representational challenge is to characterize general material histories to arbitrary accuracy with as few history variables as possible.

\subsection{Differential representations}

Differential models of inelasticity (cf., e.~g., \cite{Flugge:1975}) offer the advantage of reducing history dependence to short histories of stress and strain. Differential materials are characterized by a differential constraint of the form
\begin{equation}\label{k8exoU}
    \mbs{f}_e
    (
        \{\mbs{\epsilon}_e^{(\alpha)}(t)\}_{\alpha=0}^p, \{\mbs{\sigma}_e^{(\beta)}(t)\}_{\beta=0}^q
    )
    =
    \mbs{0} ,
\end{equation}
between the strain and its first $p$ time derivatives and stress and its first $q$ derivatives, for some material-specific function $\fb_e$ taking values in $\mathbb{R}^{m_e}$. In a time-discrete setting, the time derivatives are replaced by divided-difference formulas of the form
\begin{equation}\label{X21vSf}
    \mbs{z}_{e,k+1}^{(\alpha)}
    =
    \sum_{l=0}^\alpha
        c_{k+1,\alpha,l} \, \mbs{z}_{e,k+1-l} ,
\end{equation}
for some coefficients $\{c_{k+1,\alpha,l}\}_{l=0}^\alpha$ dependent on the choice of discrete times $\{t_{k+1-l}\}_{l=0}^\alpha$. For constant time step,
\begin{equation}
    \mbs{z}_{e,k+1}^{(\alpha)}
    =
    \frac{1}{\Delta t^\alpha}
    \sum_{l=0}^\alpha
        (-1)^l
        {\alpha \choose l}
        \mbs{z}_{e,k+1-l} ,
\end{equation}
with coefficients independent of $k+1$ as expected. Inserting these formulas into (\ref{k8exoU}), we obtain a relation of the form
\begin{equation}\label{fYCd00}
    \mbs{f}_{e}
    (
        \{\mbs{\epsilon}_{e,k+1-l}\}_{l=0}^p,
        \{\mbs{\sigma}_{e,k+1-l}\}_{l=0}^q
    )
    =
    \mbs{0} ,
\end{equation}
between the short histories of strain of length $p$ and short histories of stress of length $q$.

In this representation, the local material data sets (\ref{PS6guZ}) take the form
\begin{equation}\label{32DiI8}
    D_{e,k+1}
    =
    \{
        (\mbs{\epsilon}_{e,k+1}, \mbs{\sigma}_{e,k+1}) \, : \,
        (
            \{\mbs{\epsilon}_{e,k-l}\}_{l=0}^{p-1},
            \{\mbs{\sigma}_{e,k-l}\}_{l=0}^{q-1}
        )
    \} ,
\end{equation}
i.~e., consist of all pairs $(\mbs{\epsilon}_{e,k+1}, \mbs{\sigma}_{e,k+1})$ of stress and strain at time $t_{k+1}$ that are attainable, or known to be attainable, to the material element given the past short histories of stress and strain $(\{\mbs{\epsilon}_{e,k-l}\}_{l=0}^{p-1}, \{\mbs{\sigma}_{e,k-l}\}_{l=0}^{q-1})$ .

We note from (\ref{32DiI8}) that, for differential models, the material data set (\ref{32DiI8}) indeed depends on history through short histories of stress and strain. This parametrization is in contrast with that obtained from general representations of materials with memory, eq.~(\ref{2oKN63}), in which the history dependence of the material data set is parameterized in terms of entire, or long, histories of strain only. We thus conclude that conditioning of material data sets by means of both stress and strain histories may result in smaller parameterizations than otherwise required when only strain histories are accounted for. It may also be reasonably expected that increasing the order of differential representations (\ref{32DiI8}) should lead to increasingly accurate, and in the limit exact, representations of broad classes of materials.

\subsection{Equivalence between the internal variable and differential formalisms}
\label{YF98LU}

The correspondence between the internal variable and differential formalisms can be established as follows. For simplicity, we specifically assume internal variables of the form $\bq = \{\bq_1, \dots, \bq_N\}$, with $\bq_i \in \mathbb{R}^{m_e}$. This assumption sets the tensorial character of the internal variables to be that of a collection of internal strains. Begin by writing (\ref{e36oQ2}) as
\begin{equation}\label{BJYKd4}
    \mbs{\sigma}_e(t)
    =
    \fb_0(\mbs{\epsilon}_e(t), \bq_e(t)) .
\end{equation}
Assuming sufficient differentiability, we can differentiate this relation with respect to time and combine the result with the kinetic relations (\ref{Lg7W5y}) to obtain the identity
\begin{equation}
\begin{split}
    \dot{\mbs{\sigma}}_e(t)
    & =
    D_1\fb_0(\mbs{\epsilon}_e(t), \bq_e(t)) \dot{\mbs{\epsilon}}_e(t)
    \\ & +
    D_2\fb_0(\mbs{\epsilon}_e(t), \bq_e(t))
    D\psi_e^{-1}(- D_2F_e(\mbs{\epsilon}_e(t), \bq_e(t)))
    \\ & \equiv
    \fb_1(\mbs{\epsilon}_e(t), \dot{\mbs{\epsilon}}_e(t), \bq_e(t)) .
\end{split}
\end{equation}
Iterating this process, we obtain the system of equations
\begin{equation}\label{Vg0NTD}
    \mbs{\sigma}_e^{(\alpha)}(t)
    =
    \fb_\alpha(\{\mbs{\epsilon}_e^{(\beta)}(t)\}_{\beta=0}^\alpha, \bq_e(t)) ,
    \quad
    \alpha = 1,\dots,N ,
\end{equation}
with the functions $\fb_\alpha$ defined recursively. Assuming solvability, system (\ref{Vg0NTD}) can be solved for the internal variables to obtain a hereditary relation of the form
\begin{equation}
    \bq_e(t)
    =
    \hat{\bq}_e
    (
        \{\mbs{\epsilon}_e^{(\alpha)}(t)\}_{\alpha=0}^N,
        \{\mbs{\sigma}_e^{(\alpha)}(t)\}_{\alpha=0}^N
    ) .
\end{equation}
Inserting this relation in (\ref{BJYKd4}), we obtain the differential constraint
\begin{equation}
    \mbs{\sigma}_e(t)
    -
    \fb_0
    (
        \mbs{\epsilon}_e(t),
        \hat{\bq}_e
        (
            \{\mbs{\epsilon}_e^{(\alpha)}(t)\}_{\alpha=0}^N,
            \{\mbs{\sigma}_e^{(\alpha)}(t)\}_{\alpha=0}^N
        )
    )
    =
    0 ,
\end{equation}
which is of the general form (\ref{k8exoU}).

A similar connection can be forged directly in the time-discrete setting. Thus, iterating the propagator (\ref{iQG1K4}), we obtain the system of equations
\begin{equation}\label{IZI84d}
\begin{split}
    &
    \mbs{\sigma}_{e,k+1-l}
    = \\ &
    D_1F_e(\mbs{\epsilon}_{e,k+1-l}, \bP_e \cdots
    \bP_e(\bP_e(\bq_{k+1-N},\mbs{\epsilon}_{k+1-N}),
    \mbs{\epsilon}_{k-N},\cdots,\mbs{\epsilon}_{k+1-l})) ,
\end{split}
\end{equation}
for $l=1,\dots,N$. Assuming again solvability, the system (\ref{IZI84d}) can be solved to obtain
\begin{equation}
    \bq_{k+1-N}
    =
    \hat{\bq}_{k+1-N}
    (
        \{\mbs{\epsilon}_{e,k-l}\}_{l=0}^{N-1},
        \{\mbs{\sigma}_{e,k-l}\}_{l=0}^{N-1}
    )
\end{equation}
and
\begin{equation}
\begin{split}
    &
    \mbs{\sigma}_{e,k+1}
    = \\ &
    D_1F_e(\mbs{\epsilon}_{e,k+1},
    \bP_e \cdots
    \bP_e(\bP_e(\bq_{k+1-N},\mbs{\epsilon}_{k+1-N}),
    \mbs{\epsilon}_{k-N},\cdots,\mbs{\epsilon}_{k+1})) ,
\end{split}
\end{equation}
which supplies a time-discrete differential representation of the form (\ref{fYCd00}).

We thus conclude that internal variable and differential representations of material behavior are equivalent when the constitutive relations are sufficiently differentiable and the material behavior is stable. As already noted, within a Data-Driven framework the key conceptual advantage of the differential representation is that it relies on fundamental data only, namely, stress and strain data, and the internal variable set, if any, need not be known.

\section{Numerical examples: Viscoelasticity}
\label{IXEV5B}

We proceed to illustrate the preceding representational paradigms, and the Data-Driven schemes that they engender, by means of selected examples of application. Viscoelasticity is characterized by the smoothness of the kinetic equations and the existence of a stable equilibrium manifold. The corresponding data sets of viscoelasticity therefore lend themselves ideally to a differential representation, eqs.~(\ref{k8exoU}) and (\ref{fYCd00}).

\subsection{Example: The Standard Linear Solid}

The Standard Linear Solid, consisting of a Maxwell unit in parallel with an elastic unit, provides a simple and convenient example. The Standard Linear Solid Helmholtz free energy is
\begin{equation}\label{FOav7e}
    F_e(\mbs{\epsilon}_e, \bq_e)
    =
    \frac{1}{2}
    \mathbb{E}_0 \, \mbs{\epsilon}_e \cdot \mbs{\epsilon}_e
    +
    \frac{1}{2}
    \mathbb{E}_1
    (\mbs{\epsilon}_e - \bq_e)
    \cdot
    (\mbs{\epsilon}_e - \bq_e)
\end{equation}
where $\bq_e \in \mathbb{R}^{m_e}$ is an internal inelastic strain and $\mathbb{E}_0$ and $\mathbb{E}_1$ are moduli. The corresponding equilibrium relations (\ref{O4cXId}) are
\begin{subequations}
\begin{align}
    & \label{wrlU1l}
    \mbs{\sigma}_e(t)
    =
    D_1 F_e(\mbs{\epsilon}_e(t), \bq_e(t))
    =
    \mathbb{E}_0 \, \mbs{\epsilon}_e(t)
    +
    \mathbb{E}_1
    (\mbs{\epsilon}_e(t) - \bq_e(t))
    \\ & \label{s1lakO}
    \bp_e(t)
    =
    -
    D_2 F_e(\mbs{\epsilon}_e(t), \bq_e(t))
    =
    \mathbb{E}_1
    (\mbs{\epsilon}_e(t) - \bq_e(t)) ,
\end{align}
\end{subequations}
where $\bp_e$ is the thermodynamic driving force conjugate to $\bq_e$. Assuming linear kinetics, we further have
\begin{equation}\label{BoA7iu}
    \mathbb{E}_1
    \dot{\bq}_e(t)
    =
    \frac{\bp_e(t)}{\tau_1}
    =
    \frac{\mathbb{E}_1}{\tau_1}
    (\mbs{\epsilon}_e(t) - \bq_e(t))
\end{equation}
where $\tau_1$ is a relaxation time.

A straightforward calculation shows that the inelastic strain $\bq_e(t)$ can be eliminated from the above equations, using the time-derivative of (\ref{wrlU1l}) in addition, and that the resulting differential constraint is
\begin{equation}\label{8ZP3cV}
    \mbs{\sigma}_e(t)
    +
    \tau_1 \dot{\mbs{\sigma}}_e(t)
    -
    \mathbb{E}_0 \mbs{\epsilon}_e(t)
    -
    (\mathbb{E}_0 + \mathbb{E}_1) \tau_1 \dot{\mbs{\epsilon}}_e(t)
    =
    0 ,
\end{equation}
which is of the form (\ref{k8exoU}). A straightforward time discretization further gives
\begin{equation}\label{5jkpKD}
    \mbs{\sigma}_{e,k+1}
    +
    \tau_1
    \frac{\mbs{\sigma}_{e,k+1} - \mbs{\sigma}_{e,k}}{t_{k+1} - t_k}
    -
    \mathbb{E}_0
    \mbs{\epsilon}_{e,k+1}
    -
    (\mathbb{E}_0 + \mathbb{E}_1) \tau_1
    \frac{\mbs{\epsilon}_{e,k+1} - \mbs{\epsilon}_{e,k}}{t_{k+1} - t_k}
    =
    0 ,
\end{equation}
The corresponding differential representation (\ref{32DiI8}) of the data set is
\begin{equation}\label{E7DbKS}
    D_{e,k+1}
    =
    \{
        (\mbs{\epsilon}_{e,k+1}, \mbs{\sigma}_{e,k+1})
        \, : \,
        (\mbs{\epsilon}_{e,k}, \mbs{\sigma}_{e,k})
        \text{ and }
        (\ref{5jkpKD})
    \} ,
\end{equation}
which, for fixed $(\mbs{\epsilon}_{e,k}, \mbs{\sigma}_{e,k})$, defines a linear subspace of phase space of dimension $\mathbb{R}^{m_e}$. We conclude that first-order differential representations of the data set of the form (\ref{32DiI8}), with $p=q=1$, suffice to represent the Standard Linear Solid exactly. More generally, first-order differential representations of the form (\ref{32DiI8}) can only be expected to furnish an approximation of the actual, and unknown, material behavior.

\begin{figure}[h]
    \centering
    \includegraphics[width=0.75\linewidth]{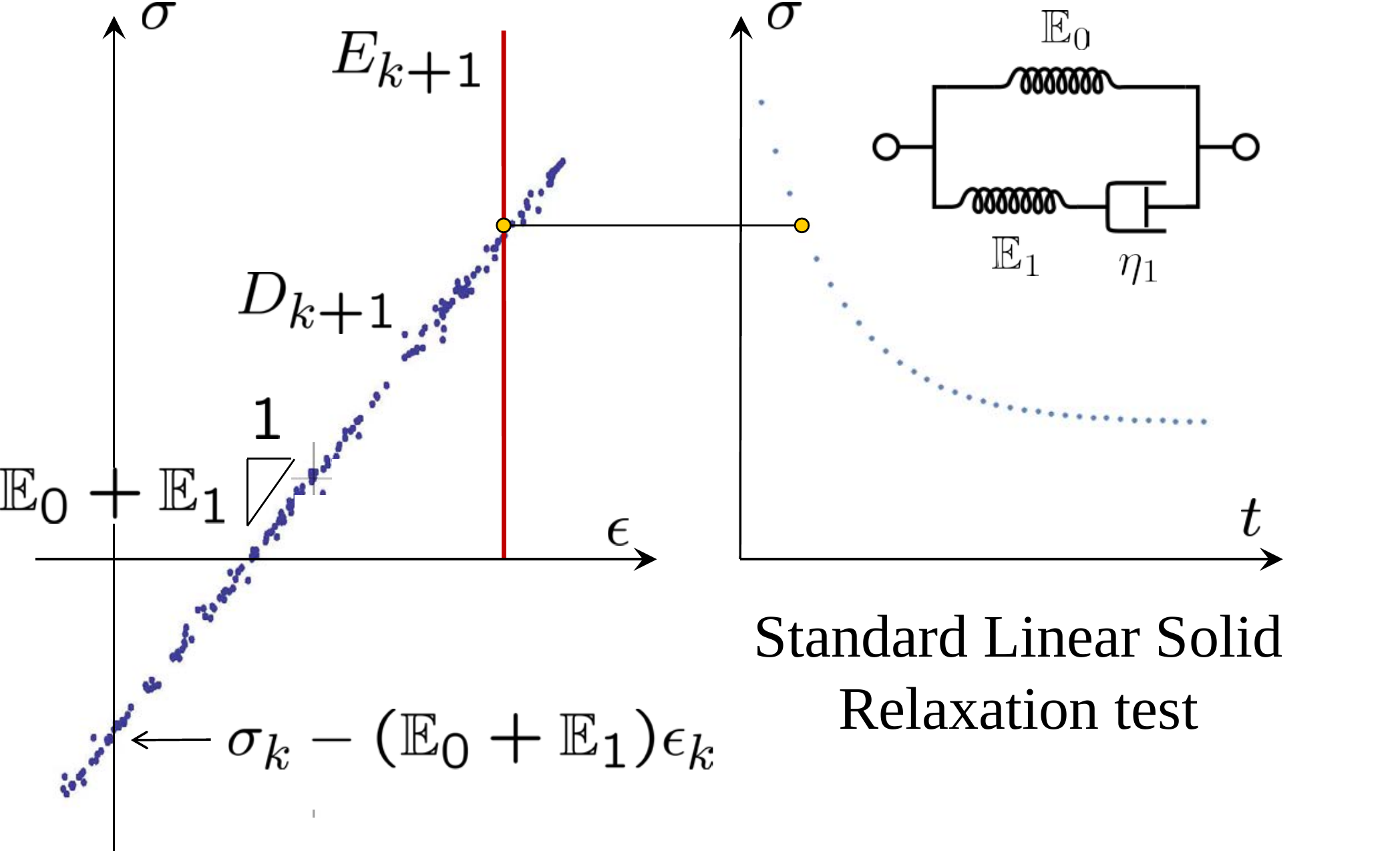}
    \caption{Schematic representation of the relaxation test for a Standard Linear Solid bar. Inlaid expressions shown in the limit $\Delta t \to 0$ for simplicity. The constraint set $E_{k+1}$, left, is fixed at a constant strain while the data set $D_{k+1}$ moves downward parallel to itself so as to trace the relaxation curve of the bar, right.} \label{7CBiWT}
\end{figure}

\subsection{Example: The relaxation test}

We illustrate the Data-Driven problem defined by the Standard Linear Solid by means of the simple example of relaxation test of a bar, Fig.~\ref{7CBiWT}. In this case, the solution consists of a single time-dependent stress and strain pair $(\epsilon(t),\sigma(t))$. The constraint set $E_{k+1}$ is then constant and simply restricts the strain to be constant and equal to a prescribed value $\bar{\epsilon}$, i.~e.,
\begin{equation}
    E_{k+1} = \{ (\epsilon, \sigma) \, : \, \epsilon = \bar{\epsilon} \} .
\end{equation}
Inserting this condition into the differential constraint (\ref{5jkpKD}), gives the relation
\begin{equation}
    {\sigma}_{k+1}
    +
    \tau_1
    \frac{{\sigma}_{k+1} - {\sigma}_{k}}{t_{k+1} - t_k}
    -
    \mathbb{E}_0
    \bar{{\epsilon}}
    =
    0 .
\end{equation}
A straightforward calculation gives the Data-Driven solution as
\begin{equation}\label{NYRK94}
\begin{split}
    \epsilon_k
    =
    \bar{\epsilon},
    \quad
    {\sigma}_k
    & =
    \mathbb{E}_1\bar{\epsilon}
    \left(\frac{\tau_1}{\Delta t + \tau_1}\right)^k
    \\ & +
    \left[
        \sum_{n=1}^{k-1}
        \left(
            \frac{\tau_1}{\Delta t + \tau_1}
        \right)^n
        \frac{\Delta t}{\Delta t + \tau_1}
        +
        \left(
            \frac{\tau_1}{\Delta t + \tau_1}
        \right)^k
    \right]
    \mathbb{E}_0 \bar{\epsilon} ,
\end{split}
\end{equation}
where we assume $t_{k+1} - t_k = \Delta t = $ constant, for simplicity. Inserting (\ref{NYRK94}) into (\ref{5jkpKD}) defines the data set $D_{k+1}$ as a line in phase space of slope approximately equal to $\mathbb{E}_0 + \mathbb{E}_1$ intersecting the stress axis at approximately $\sigma_k - (\mathbb{E}_0 + \mathbb{E}_1) \epsilon_k$.

Thus, the initial material data set $D_0$ is a line of slope roughly $\mathbb{E}_0 + \mathbb{E}_1$ through the origin that intersects the constraint set $E_0$ at $\sigma_0 = (\mathbb{E}_0 + \mathbb{E}_1) \bar{\epsilon}$, which is the instantaneous response of the solid. Subsequent material data sets $D_{k+1}$ translate downwards in phase space and their intersection with the constraint set $E_{k+1}$ traces the relaxation curve of the bar. More general Data-Driven solutions can be obtained if the material data set $D_{k+1}$ is allowed to be a point set, e.~g., approximating the Standard Linear Solid data set just described. In this case, the Data-Driven solution is the point in the constraint set $E_{k+1}$ closest to the material data set $D_{k+1}$. With the passage of time, these points again trace a Data-Driven relaxation curve of the bar, Fig.~\ref{7CBiWT}.

\begin{figure}[h]
  \centering
  \includegraphics[width=0.65\linewidth]{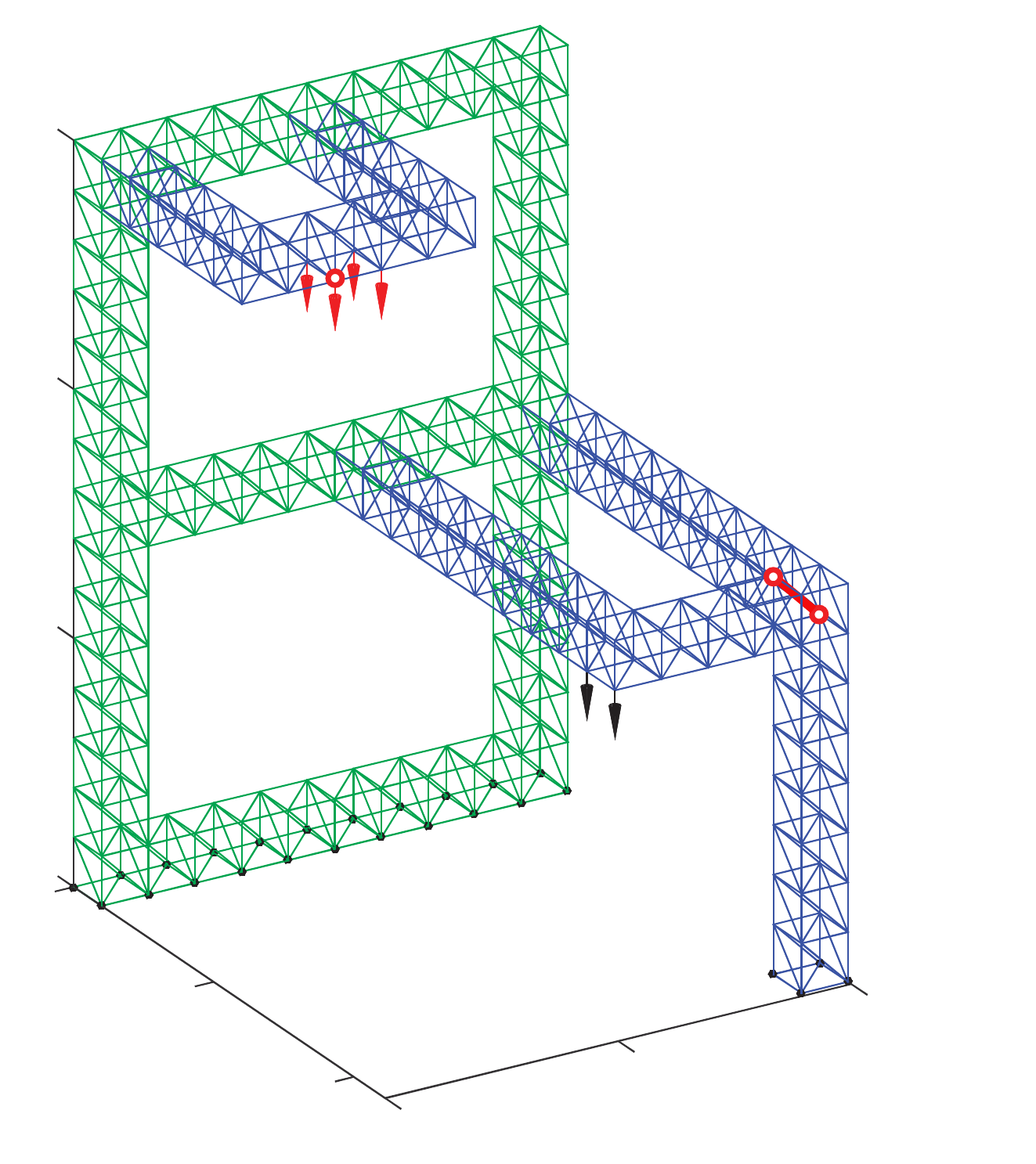}
  \caption{Truss Geometry, load points, and output locations. Red arrows represent applied loads, black arrows prescribed displacements. Nodes highlighted in black are fixed. Vertical displacements are output and monitored at the node highlighted in red. Member forces are output and monitored at the bar highlighted in red. }
  \label{4NZE8X}
\end{figure}

\subsection{Convergence analysis: Truss structures}
\label{BuKeb1}

We demonstrate the convergence properties of Data-Driven viscoelasticity with the aid of the three-dimensional truss structure shown in Fig.~\ref{4NZE8X}. The geometry of the truss, which comprises 1,246 bars, the boundary conditions and the applied loads are also shown in Fig.~\ref{4NZE8X}. The loads are linearly ramped up to $t=10$, subsequently held constant up to $t=50$, linearly ramped back to zero at $t=60$, and held again constant up to $t=100$. The data sets are generated on the fly by randomizing the Standard Linear Solid data set (\ref{E7DbKS}). The data points are assumed to be uniformly distributed within a band of width $\Delta \epsilon_{e,k+1} = 0.030$. A typical local material data set is shown in Fig.~\ref{7CBiWT}. The resulting material data sets converge uniformly to the Standard Linear Solid graph in the sense defined in \cite{Conti:2018}. The parameters of the reference Standard Linear Solid used in calculations are $\mathbb{E}_0 = 75,000$, $\mathbb{E}_1 = 100,000$ and $\tau_1 = 5$. In addition, a constant time step $\Delta t = 1$ is used in all calculations.

\begin{figure}[h]
    \centering
    \begin{subfigure}{0.49\textwidth}
    \includegraphics[width=0.99\linewidth]{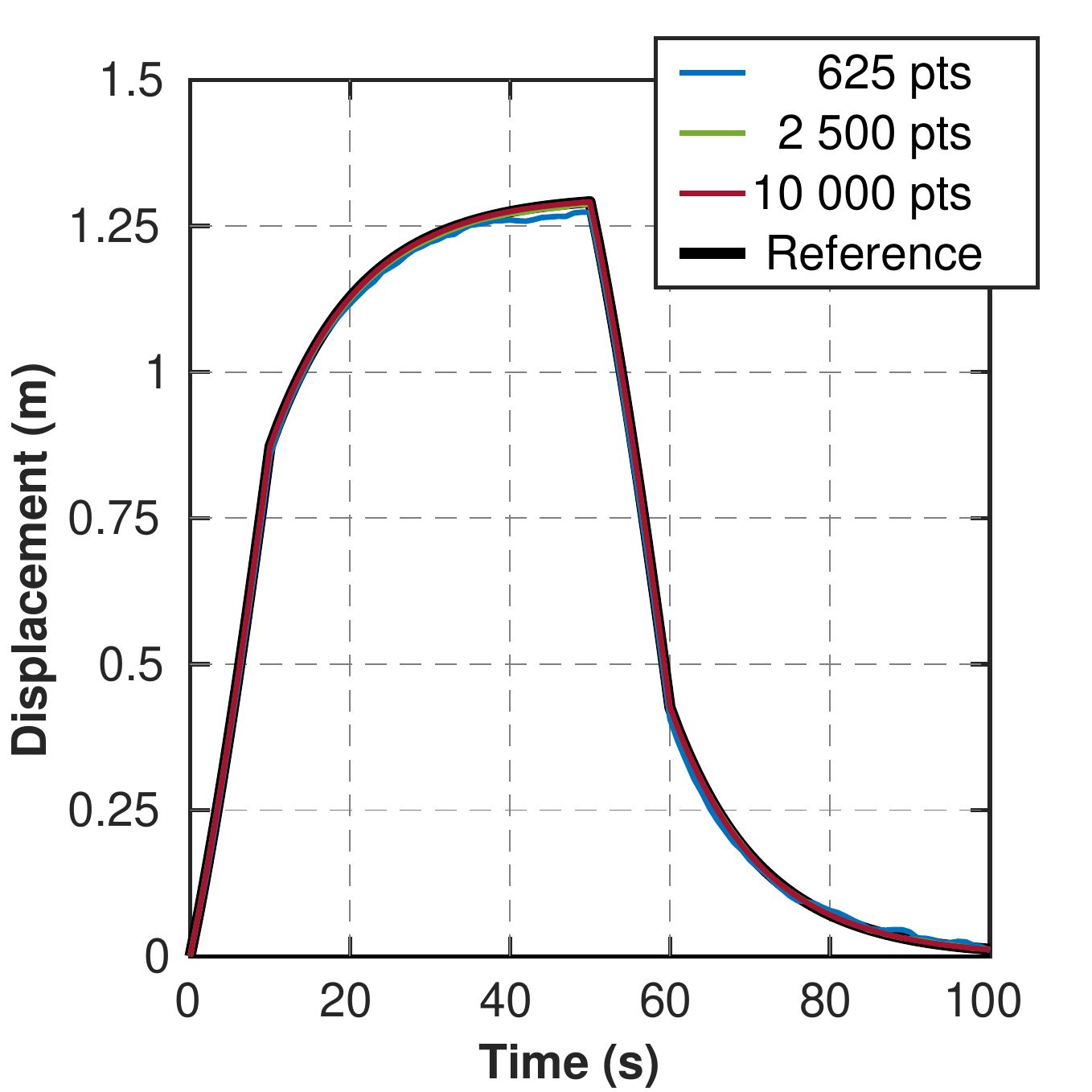}
    \caption{}
    \label{w8e4TM}
    \end{subfigure}
    \begin{subfigure}{0.49\textwidth}
    \includegraphics[width=0.99\linewidth]{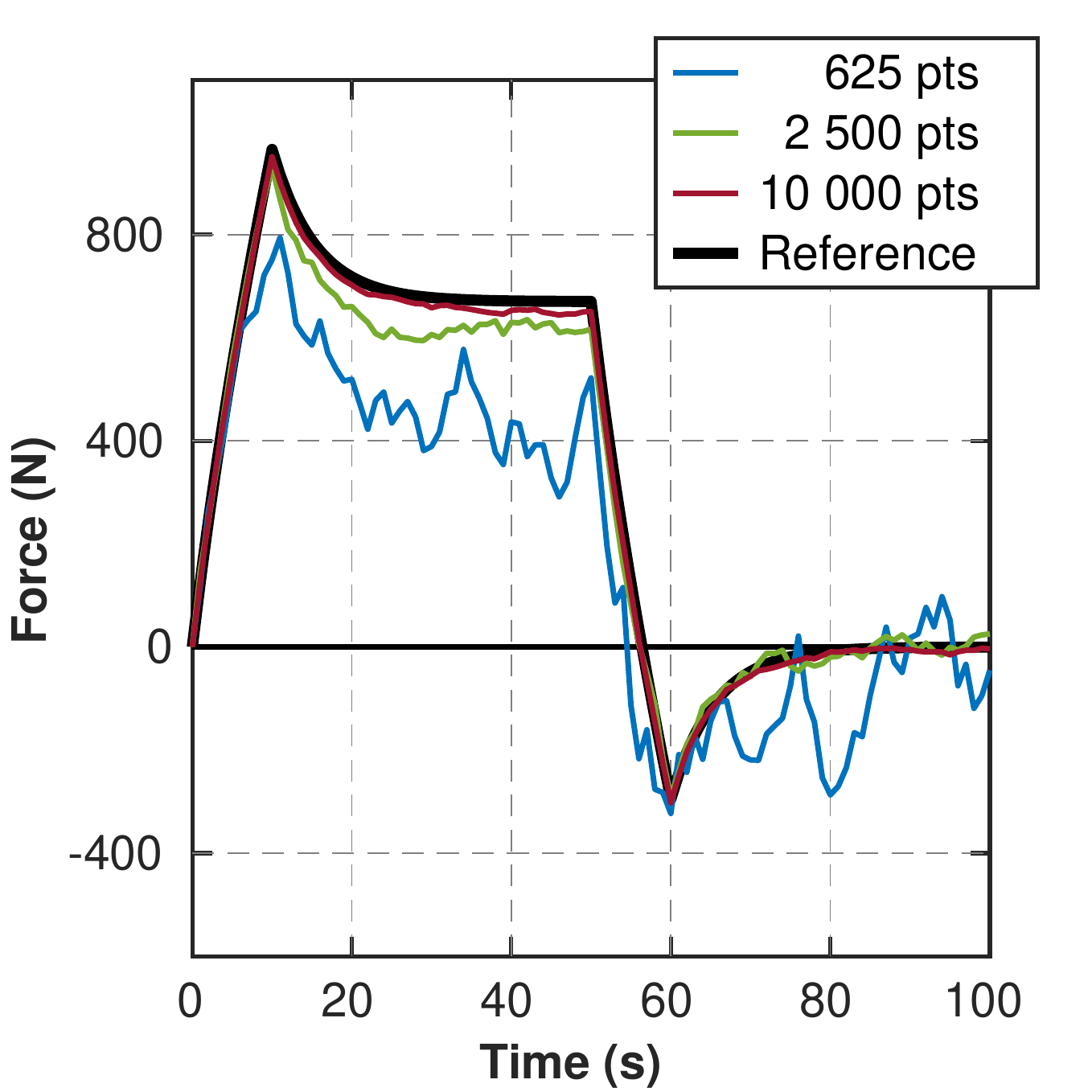}
    \caption{}
    \label{lF01xo}
    \end{subfigure}
    \caption{Viscoelastic truss problem. Time-history comparison for data solver at various data resolutions for a) deflections at a degree of freedom with an applied force and b) axial forces in output bar.}
    \label{Z4K9kb}
\end{figure}

\begin{figure}[h]
    \centering
    \includegraphics[width=0.65\linewidth]{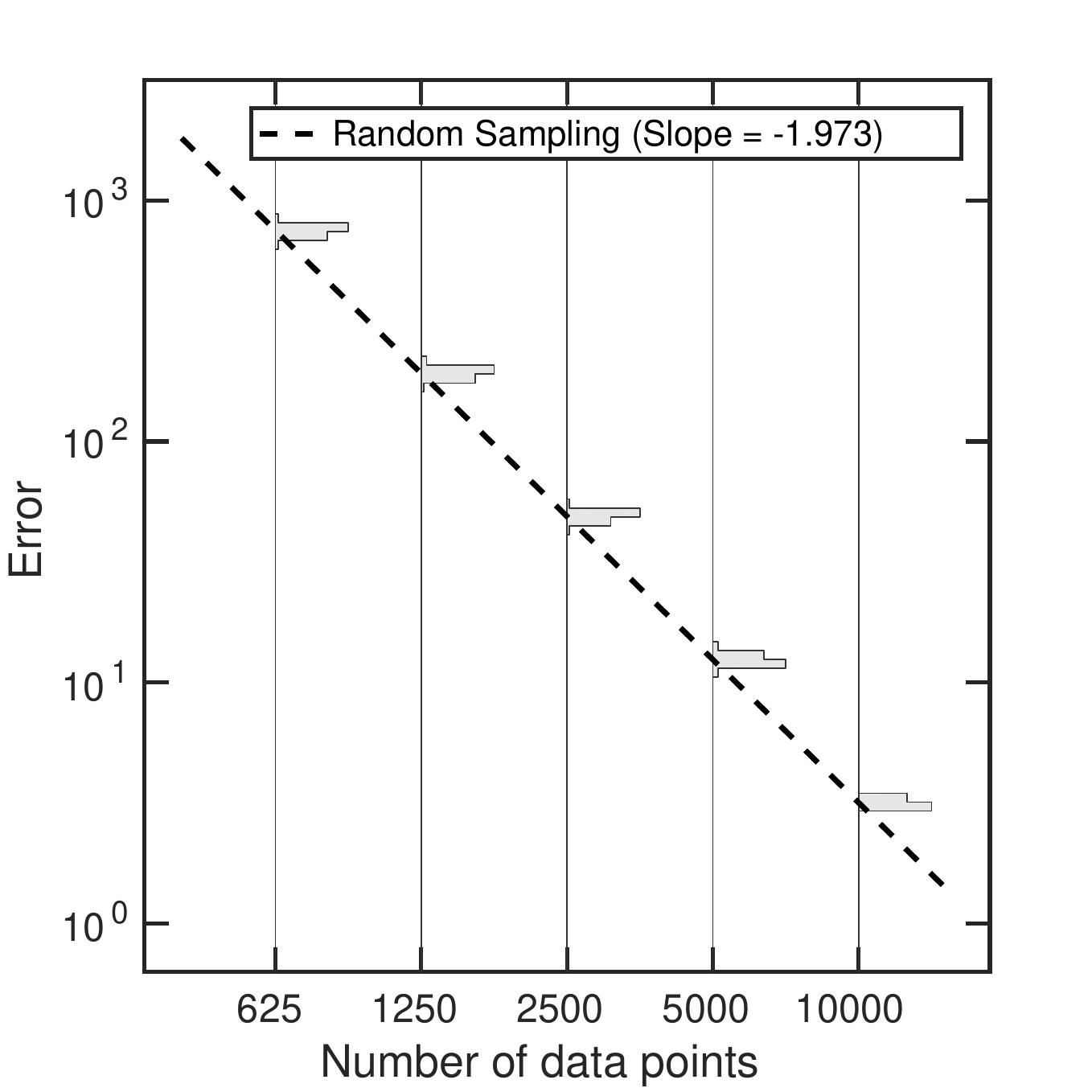}
    \caption{Viscoelastic truss problem. Data convergence of the Data Driven viscoelastic problem to the reference Standard Linear Solid solution.}
    \label{m5VKwr}
\end{figure}

Fig.~\ref{w8e4TM} depicts displacement histories at the output node shown in Fig.~\ref{4NZE8X} and Fig.~\ref{lF01xo} shows the history of the resultant of the reaction forces at the kinematically constraint nodes, cf.~Fig.~\ref{4NZE8X}. The convergence of the time histories towards the solution of the reference Standard Linear Solid with increasing number of materials data points is evident in the figures. The rate of convergence can be monitored by means of the weighted $\ell^2$ error
\begin{equation}\label{c558Xj}
    \text{Error}
    =
    \left(
        \sum_{k=0}^{T-1}
        |z_{k+1} - z_{k+1}^{\text{ref}}|^2 \,
        {\rm e}^{-t_{k+1}/\tau_1} (t_{k+1} - t_k)
    \right)^{1/2}
\end{equation}
where $T$ is the number of time steps, $| \cdot |$ is as in (\ref{9oakLa}) and $z_{e,k}^{\text{ref}} = (\epsilon_{e,k}^{\text{ref}}, \sigma_{e,k}^{\text{ref}} )$ is the solution for the reference Standard Linear Solid. Weighted norms such as (\ref{c558Xj}) arise naturally in the analysis of viscoelastic problems (cf., e.~g., \cite{Lions:1972}). Compiling statistics over $50$ independent runs, i.~e., with different randomizations of the data set, we arrive at the convergence plot shown in Fig.~\ref{m5VKwr}. Remarkably, the computed rate of convergence is quadratic, or twice the linear rate of convergence characteristic of elastic problems \cite{Kirchdoerfer:2016}.

\section{Numerical examples: Plasticity}
\label{X0aZjS}

\begin{figure}[h]
    \centering
    \includegraphics[width=0.75\linewidth]{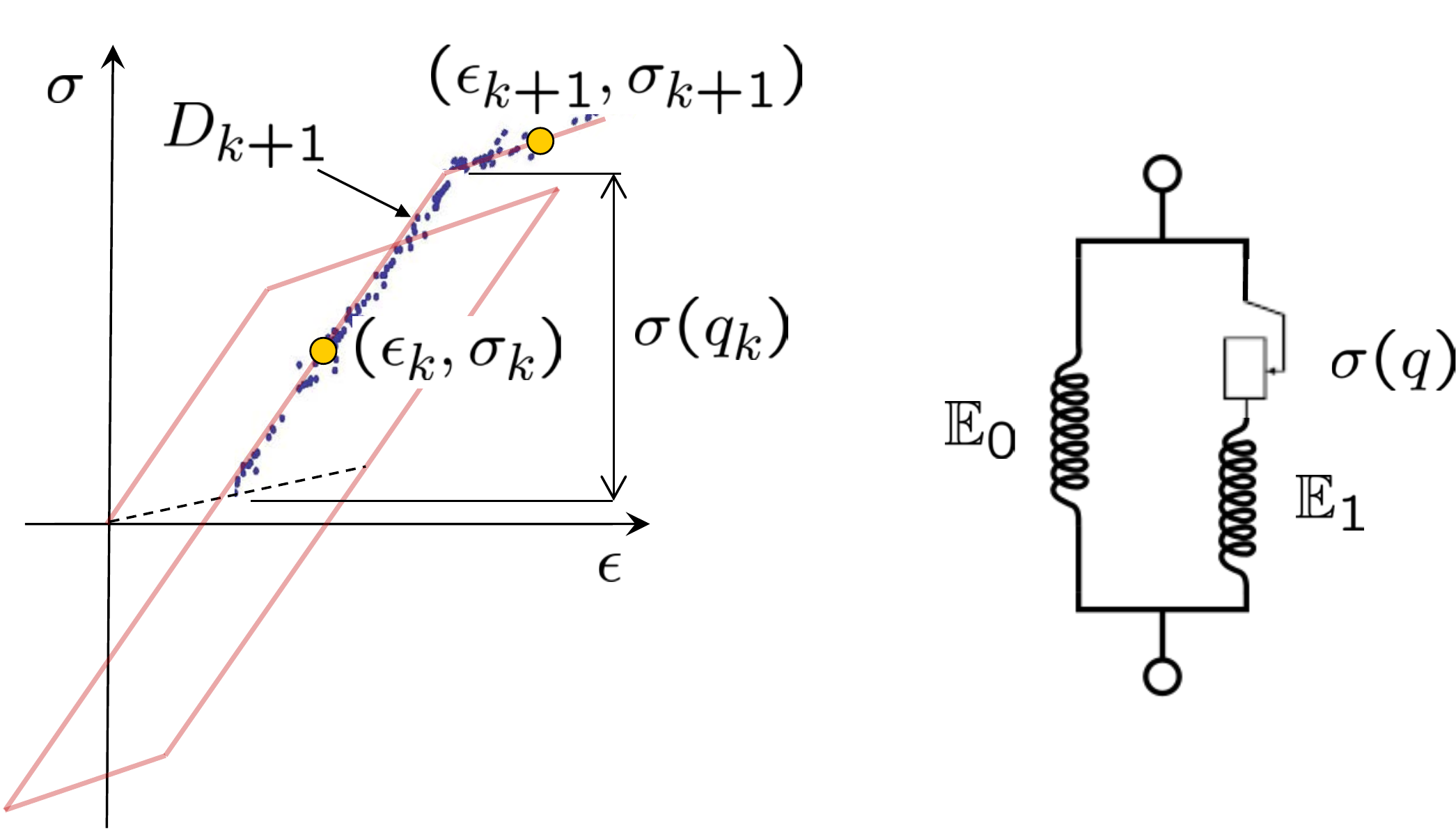}
    \caption{Schematic representation of the evolution of a typical data set for the isotropic-kinematic linear-hardening solid, left. Rheological model consisting of two elastic elements and a hardening slider, right.}
    \label{1ZESqU}
\end{figure}

Plasticity (cf., e.~g., \cite{Lubliner:1990}) supplies an example of a class of material data sets that are not amenable to a strict differential representation and require the use of history variables in addition.

\subsection{Example: The isotropic-kinematic linear-hardening solid}

We illustrate this class of materials by means of the simple isotropic-kinematic linear-hardening solid, Fig.~\ref{1ZESqU}. In this case, the free energy is of the form
\begin{equation}
    F_e(\mbs{\epsilon}_e, \bq_e, q_e)
    =
    \frac{1}{2}
    \mathbb{E}_0 \, \mbs{\epsilon}_e \cdot \mbs{\epsilon}_e
    +
    \frac{1}{2}
    \mathbb{E}_1
    (\mbs{\epsilon}_e - \bq_e)
    \cdot
    (\mbs{\epsilon}_e - \bq_e)
    +
    W_e(q_e)
\end{equation}
where $\bq_e \in \mathbb{R}^{m_e}$ is a internal inelastic strain, $q_{e}$ is an effective accumulated plastic strain, $W_e$ is a stored energy of cold work and $\mathbb{E}_0$ and $\mathbb{E}_1$ are moduli. The equilibrium relations (\ref{O4cXId}) evaluate to
\begin{subequations}
\begin{align}
    & \label{B85dgt}
    \mbs{\sigma}_e(t)
    =
    \mathbb{E}_0 \mbs{\epsilon}_e(t)
    +
    \mathbb{E}_1 (\mbs{\epsilon}_e(t) - \bq_e(t))
    \\ &
    \bp_e(t)
    =
    \mathbb{E}_1 (\mbs{\epsilon}_e(t) - \bq_e(t))
    \\ &
    - p_e(t)
    =
    W_e'(q_e)
    \equiv
    \sigma_e(q_e) ,
\end{align}
\end{subequations}
where $\sigma_e(q_e)$ is the yield stress. For the rate-independent solid, the dual kinetic potential is of the form
\begin{equation}\label{NDKmA2}
    \psi^*(\bp_e , p_e)
    =
    \left\{
        \begin{array}{ll}
            0, & \text{if }
            f(\bp_e , p_e)  \leq 0 , \\
            +\infty, & \text{otherwise} ,
        \end{array}
    \right.
\end{equation}
for some convex yield function $f(\bp_e , p_e)$, i.~e., $\psi^*$ vanishes within the elastic domain $f(\bp_e , p_e) \leq 0$ and equals $+\infty$ elsewhere in driving-force space. We note that $\psi^*(\bp_e , p_e)$ is not differentiable and, therefore, the corresponding kinetic relations
\begin{equation}\label{Ie6NO0}
    (\dot{\bq}_e(t), \dot{q}_e(t))
    \in
    \partial
    \psi^*(\bp_e(t) , p_e(t))
\end{equation}
are set-valued and must be understood in the sense of subdifferentials \cite{Rockafellar:1970}. Equivalently, the kinetic relations (\ref{Ie6NO0}) can be expressed in terms of Drucker's principle of maximum dissipation
\begin{equation}\label{8DEmAC}
    \max_{(\bp_e(t) , p_e(t))}
    \big\{
        \bp_e(t) \cdot \dot{\bq}_e(t)
        +
        p_e(t) \dot{q}_e(t)
        -
        \psi^*(\bp_e(t) , p_e(t))
    \big\} ,
\end{equation}
where the rates $(\dot{\bq}_e(t),\dot{q}_e(t))$ are regarded as given. In view of (\ref{NDKmA2}), (\ref{8DEmAC}) is in turn equivalent to
\begin{equation}\label{5VnV1o}
    \max_{(\bp_e(t) , p_e(t))}
    \big\{
        \bp_e(t) \cdot \dot{\bq}_e(t)
        +
        p_e(t) \dot{q}_e(t)
        \, : \,
        f(\bp_e(t) , p_e(t)) \leq 0
    \big\} ,
\end{equation}
which defines a standard convex-optimization problem \cite{Rockafellar:1970}. Introducing a Lagrange multiplier $\lambda_e(t)$, the corresponding Euler-Lagrange equations are
\begin{subequations}
\begin{align}
    &
    \dot{\bq}_e(t)
    =
    \lambda_e(t)
    \frac{\partial f}{\partial \bq_e}(\bp_e(t) , p_e(t)) ,
    \\ &
    \dot{q}_e(t)
    =
    \lambda_e(t)
    \frac{\partial f}{\partial q_e}(\bp_e(t) , p_e(t)) ,
\end{align}
\end{subequations}
subject to the Kuhn-Tucker conditions
\begin{equation}
    f(\bp_e(t) , p_e(t)) \leq 0 ,
    \quad
    \lambda_e(t) \geq 0 ,
    \quad
    f(\bp_e(t) , p_e(t)) \lambda_e(t) = 0 ,
\end{equation}
which encode the yielding and loading-unloading conditions. A fully-implicit  discretization of (\ref{5VnV1o}) gives the time-discrete maximum dissipation principle
\begin{equation}
    \max_{(\bp_{e,k+1} , p_{e,k+1})}
    \big\{
        \bp_{e,k+1} \cdot
        (\bq_{e,k+1} - \bq_{e,k})
        +
        p_{e,k+1}
        (q_{e,k+1} - q_{e,k})
        \, : \,
        f(\bp_{e,k+1} , p_{e,k+1}) \leq 0
    \big\} ,
\end{equation}
where $(\bq_{e,k+1}, q_{e,k+1})$ are regarded as given. The corresponding Euler-Lagrange equations are
\begin{subequations}\label{9NntA7}
\begin{align}
    & \label{D7BKjG}
    \bq_{e,k+1} - \bq_{e,k}
    =
    \lambda_{e,k+1}
    \frac{\partial f}{\partial \bq_e}(\bp_{e,k+1} , p_{e,k+1})
    \\ & \label{0mDLWD}
    q_{e,k+1} - q_{e,k}
    =
    \lambda_{e,k+1}
    \frac{\partial f}{\partial q_e}(\bp_{e,k+1} , p_{e,k+1})
\end{align}
\end{subequations}
subject to the Kuhn-Tucker loading-unloading conditions
\begin{equation}
    f(\bp_{e,k+1} , p_{e,k+1}) \leq 0 ,
    \quad
    \lambda_{e,k+1} \geq 0 ,
    \quad
    f(\bp_{e,k+1} , p_{e,k+1}) \lambda_{e,k+1} = 0 .
\end{equation}
These equations are closed by the time-discrete equilibrium relations
\begin{subequations}\label{9r6HVL}
\begin{align}
    & \label{F7BThA}
    \mbs{\sigma}_{e,k+1}
    =
    \mathbb{E}_0 \mbs{\epsilon}_{e,k+1}
    +
    \mathbb{E}_1 (\mbs{\epsilon}_{e,k+1} - \bq_{e,k+1})
    \\ &
    \bp_e(t)
    =
    \mathbb{E}_1 (\mbs{\epsilon}_{e,k+1} - \bq_{e,k+1})
    \\ &
    -
    p_{e,k+1}
    =
    W_e'(q_{e,k+1})
    \equiv
    \sigma_e(q_{e,k+1}) ,
\end{align}
\end{subequations}
and jointly define a convex problem for $(\mbs{\sigma}_{e,k+1}, \bq_{e,k+1}, q_{e,k+1})$ given $\mbs{\epsilon}_{e,k+1}$ and $(\mbs{\sigma}_{e,k}, \bq_{e,k}, q_{e,k})$. A solution of this problem can be conveniently obtained by means of an elastic predictor-plastic corrector split \cite{Ortiz:1999, Vladimirov:2008}.

We note that the material data set $D_{e,k+1}$ of points $(\mbs{\epsilon}_{e,k+1}, \mbs{\sigma}_{e,k+1})$ attainable at time $t_{k+1}$ is fully characterized by $(\mbs{\epsilon}_{e,k}, \mbs{\sigma}_{e,k})$ and $q_{e,k}$. The dependence of $D_{e,k+1}$ on $(\mbs{\epsilon}_{e,k}, \mbs{\sigma}_{e,k})$ is consistent with a differential representation. However, the additional dependence on $q_{e,k}$ is typical of a history variable representation. Indeed, the history-variable character of $q_{e,k}$ can be revealed as follows. Taking a convenient seminorm $| \cdot |$ of (\ref{D7BKjG}) and eliminating $\lambda_{e,k+1}$ together with (\ref{0mDLWD}), we obtain
\begin{equation}\label{2B0VSn}
    q_{e,k+1} - q_{e,k}
    =
    \frac{|\bq_{e,k+1} - \bq_{e,k}|}{|\partial f/\partial \bq_e(\bp_{e,k+1} , p_{e,k+1})|}
    \frac{\partial f}{\partial q_e}(\bp_{e,k+1} , p_{e,k+1}) .
\end{equation}
At this point, we note that the choice of yield function $f(\bp_{e} , p_{e})$ is arbitrary up to scaling by positive functions, since that operation leaves the elastic domain invariant. Therefore, we may choose a normalization of $f(\bp_{e} , p_{e})$ such that
\begin{equation}
    \frac
    {
        \partial f/\partial q_e(\bp_{e} , p_{e})
    }
    {
        | \partial f/\partial \bq_e(\bp_{e} , p_{e}) |
    }
    =
    1 .
\end{equation}
With this normalization, (\ref{2B0VSn}) reduces to
\begin{equation}\label{il3XRm}
    q_{e,k+1} - q_{e,k}
    =
    |\bq_{e,k+1} - \bq_{e,k}|
\end{equation}
From (\ref{F7BThA}), we additionally have
\begin{equation}
    \bq_{e,k+1} - \bq_{e,k}
    =
    \mathbb{E}_1^{-1}
    \big(
        ( \mathbb{E}_0 + \mathbb{E}_1 )
        (\mbs{\epsilon}_{e,k+1} - \mbs{\epsilon}_{e,k})
        -
        \mbs{\sigma}_{e,k+1} + \mbs{\sigma}_{e,k}
    \big) ,
\end{equation}
which, inserted into (\ref{il3XRm}), further gives the incremental relation
\begin{equation}
    q_{e,k+1} - q_{e,k}
    =
    \big|
        \mathbb{E}_1^{-1}
        \big(
            ( \mathbb{E}_0 + \mathbb{E}_1 )
            (\mbs{\epsilon}_{e,k+1} - \mbs{\epsilon}_{e,k})
            -
            \mbs{\sigma}_{e,k+1} + \mbs{\sigma}_{e,k}
        \big)
    \big| .
\end{equation}
Finally, summing over the history of the material, we obtain the relation
\begin{equation}\label{IH8SXh}
    q_{e,k}
    =
    \sum_{h \leq k}
    \big|
        \mathbb{E}_1^{-1}
        \big(
            ( \mathbb{E}_0 + \mathbb{E}_1 )
            (\mbs{\epsilon}_{e,h} - \mbs{\epsilon}_{e,h-1})
            -
            \mbs{\sigma}_{e,h} + \mbs{\sigma}_{e,h-1}
        \big)
    \big| ,
\end{equation}
which is of the form (\ref{RTL6EL}).

It follows from the preceding analysis that the material data set of an isotropic-kinematic plastic solid admits the mixed differential-hereditary representation
\begin{equation}\label{qo5b8V}
    D_{e,k+1}
    =
    \{
        (\mbs{\epsilon}_{e,k+1}, \mbs{\sigma}_{e,k+1})
        \, : \,
        (\mbs{\epsilon}_{e,k}, \mbs{\sigma}_{e,k}, q_{e,k}),
        \text{ and }
        (\ref{9NntA7}-\ref{9r6HVL})
    \} ,
\end{equation}
which, for fixed $(\mbs{\epsilon}_{e,k}, \mbs{\sigma}_{e,k}, q_{e,k})$, defines a linear subspace of phase space of dimension $\mathbb{R}^{m_e}$. Again, from a Data-Driven perspective, the right interpretation of this result is that a mixed differential-hereditary of the form (\ref{qo5b8V}) suffices to represent the isotropic-kinematic plastic solid exactly. However, for general plastic solids the history variable (\ref{IH8SXh}) represents an {\sl ad hoc} choice intended to record partial information about the history of the material. Then, representations of the form (\ref{qo5b8V}), with $D_{e,k+1}$ consisting of points $(\mbs{\epsilon}_{e,k+1}, \mbs{\sigma}_{e,k+1})$ in phase space known to be attainable from initial conditions $(\mbs{\epsilon}_{e,k}, \mbs{\sigma}_{e,k}, q_{e,k})$, cf.~Fig.~\ref{1ZESqU}, can only be expected to furnish an approximation of the actual material behavior.

\subsection{Convergence analysis: Truss structures}

We again demonstrate the convergence properties of Data-Driven plasticity with the aid of the three-dimensional truss structure shown in Fig.~\ref{4NZE8X}. The boundary conditions are as in the viscoelastic calculations of Section~\ref{BuKeb1}. The loads are linearly ramped up linearly from $0$ to $0.8$ at $t=20$, ramped down to $-0.9$ at $t=60$ and finally ramped up again to $1.0$ at $t=100$. The data sets are generated on the fly by randomizing the isotropic-kinematic linear-hardening data set (\ref{qo5b8V}). The data points are assumed to be uniformly distributed within a band of width $\Delta \epsilon_{e,k+1} = 0.04$. A typical local material data set is shown in Fig.~\ref{1ZESqU}. The resulting material data sets converge uniformly to the material data sets of isotropic-kinematic hardening solid, cf.~\cite{Conti:2018}. The parameters of the reference isotropic-kinematic hardening solid used in calculations are $\mathbb{E}_0 = 10,000$, $\mathbb{E}_1 = 100,000$ and initial yield stress $\sigma_1 = 500$. Finally, a constant time step $\Delta t = 1$ is used in all calculations.

\begin{figure}[h]
    \centering
    \begin{subfigure}{0.49\textwidth}
    \includegraphics[width=0.99\linewidth]{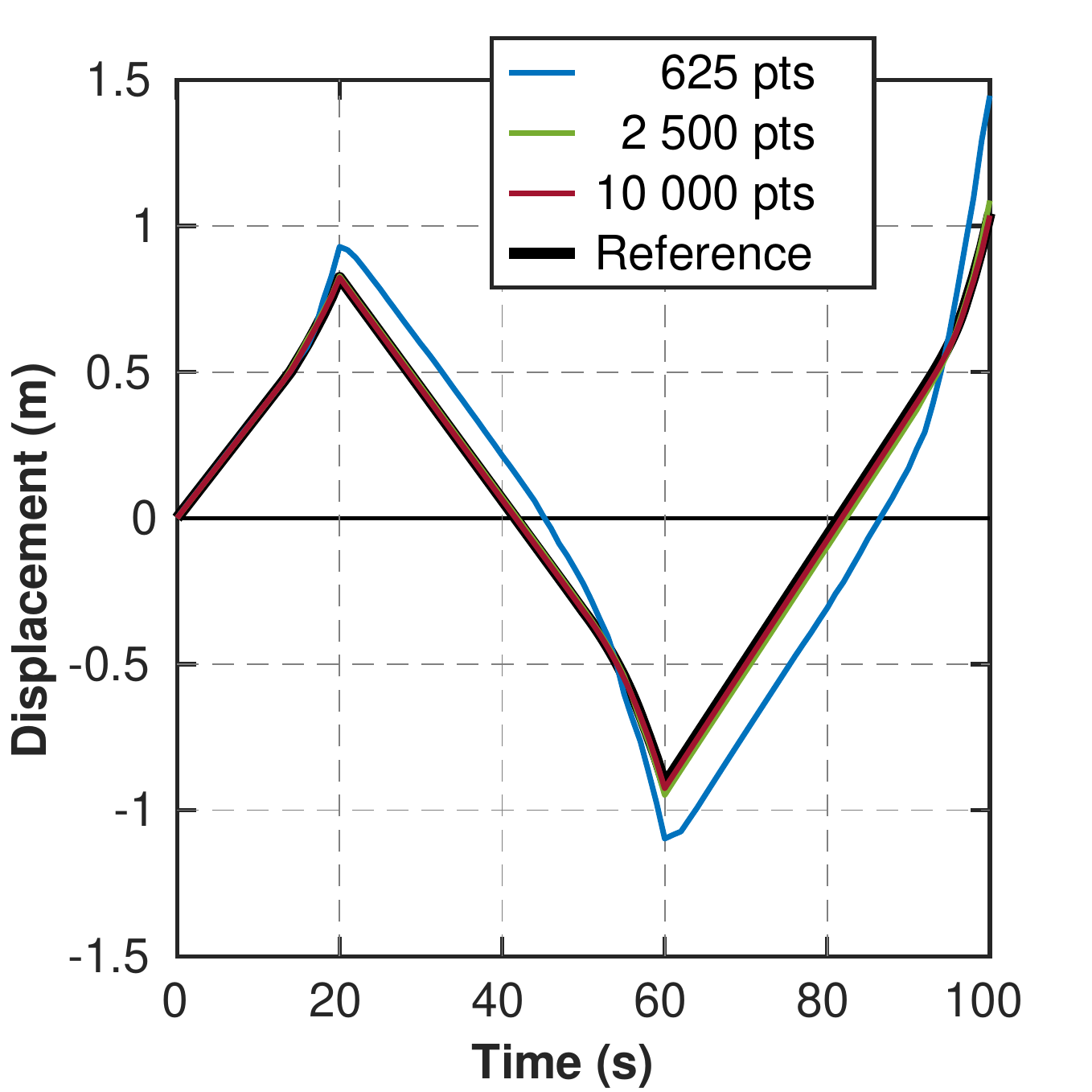}
    \caption{}
    \label{Z19SwM}
    \end{subfigure}
    \begin{subfigure}{0.49\textwidth}
    \includegraphics[width=0.99\linewidth]{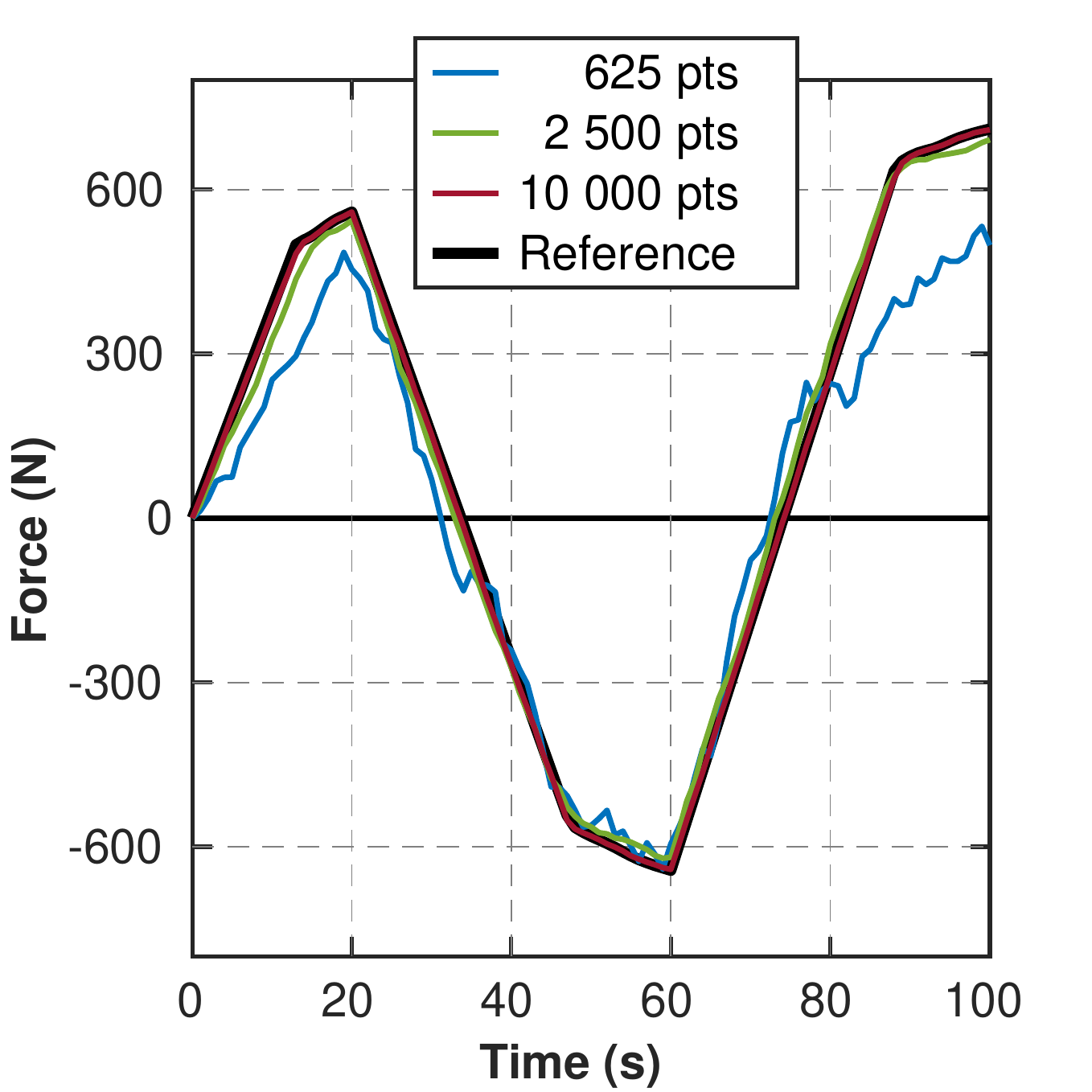}
    \caption{}
    \label{7EDnGV}
    \end{subfigure}
    \caption{Plastic truss problem. Time-history comparison for data solver at various data resolutions for a) deflections at a degree of freedom with an applied force and b) axial forces measured in output bar.}
    \label{2HAVL2}
\end{figure}

\begin{figure}[h]
    \centering
    \includegraphics[width=0.65\linewidth]{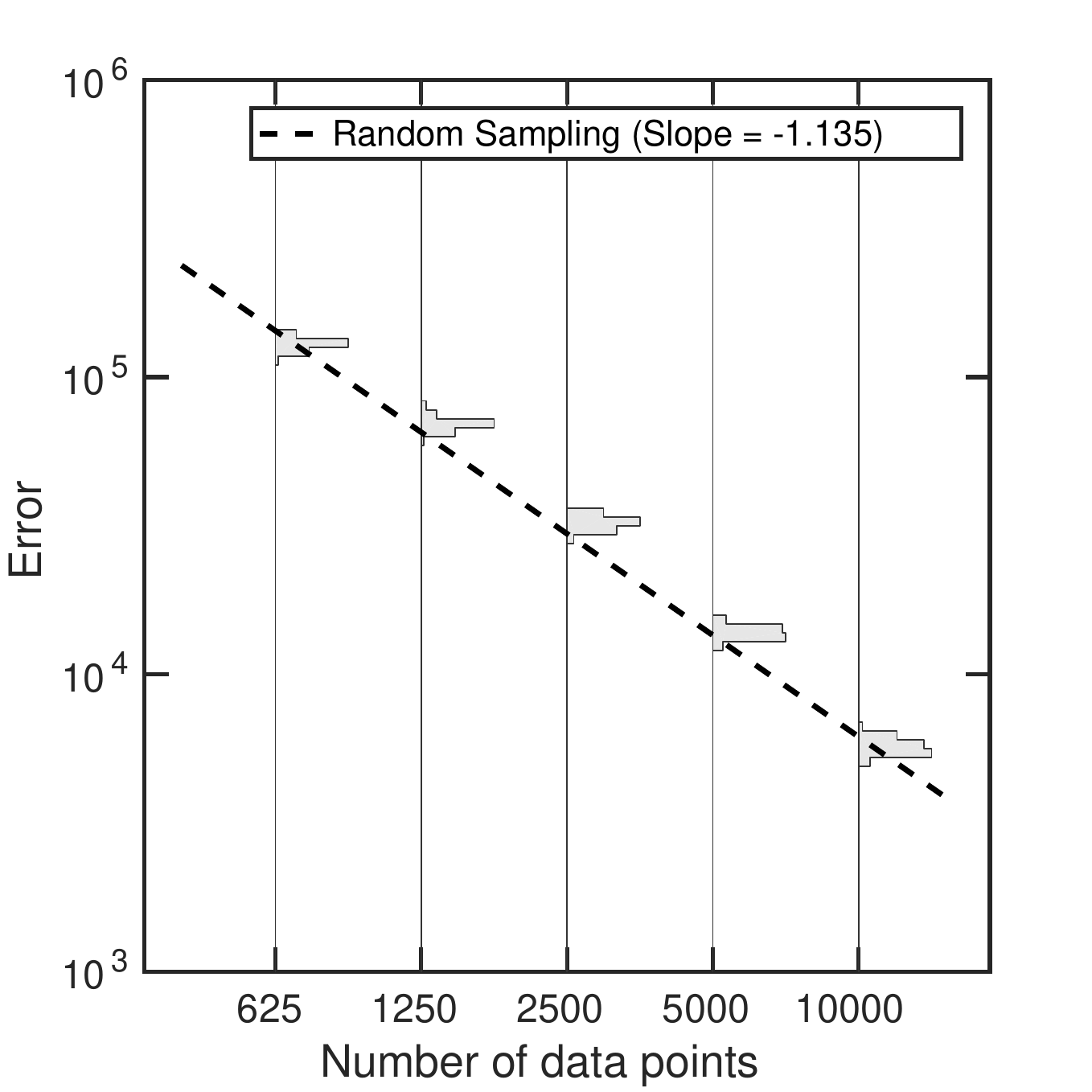}
    \caption{Plastic truss problem. Data convergence of the Data Driven viscoelastic problem to the reference isotropic-kinematic hardening solution.}
    \label{3DFGSu}
\end{figure}

Fig.~\ref{Z19SwM} depicts displacement histories at the output node shown in Fig.~\ref{4NZE8X} and Fig.~\ref{7EDnGV} shows the history of the resultant of the reaction forces at the kinematically constraint nodes, cf.~Fig.~\ref{4NZE8X}. The convergence of the time histories towards the solution of the reference isotropic-kinematic hardening solid with increasing number of materials data points is evident in the figures. The rate of convergence can be monitored by means of the rate-independent error
\begin{equation}\label{sZRRM4}
    \text{Error}
    =
    \sum_{k=0}^{T-1}
    | (\bz_{k+1} - \bz_k) - (\bz_{k+1}^{\text{ref}} - \bz_k^{\text{ref}}) |
\end{equation}
where $T$ is the number of time steps, $| \cdot |$ is as in (\ref{9oakLa}) and $z_{e+k}^{\text{ref}} = (\epsilon_{e+k}^{\text{ref}}, \sigma_{e+k}^{\text{ref}} )$ is the solution for the reference elastic-plastic Solid. Bounded-variation norms such as (\ref{sZRRM4}) arise naturally in the analysis of plasticity problems (cf., e.~g., \cite{Mielke:2008}). Compiling statistics over $50$ independent runs, we arrive at the convergence plot shown in Fig.~\ref{3DFGSu}. The computed rate of convergence is roughly linear, which coincides with the linear rate of convergence characteristic of elastic problems \cite{Kirchdoerfer:2016}.

\section{Summary and concluding remarks}
\label{y1HJx6}

We have extended the Data-Driven formulation of problems in elasticity of Kirchdoerfer and Ortiz \cite{Kirchdoerfer:2016} to inelasticity. This extension differs fundamentally from Data-Driven problems in elasticity in that the material data set evolves in time as a consequence of the history dependence of the material. Therefore, the central issue of Data-Driven inelasticity concerns the practical representation of evolving, history-dependent material data sets. In this regard, we have investigated three representational paradigms: i) {\sl materials with memory}, i.~e., conditioning the material data set to the past history of deformation; ii) {\sl differential materials}, i.~e., conditioning the material data set to short histories of stress and strain; and iii) {\sl history variables}, i.~e., conditioning the material data set to {\sl ad hoc} variables encoding partial information about the history of stress and strain. We have also considered combinations of these three paradigms thereof. We find that many classical models of viscoelasticity and plasticity can be represented by means of material data sets of the differential and/or history variable type. Evidently, such representations only afford approximations of actual, often complex, material behavior. The central approximation question therefore concerns the formulation of material set representations capable of accounting increasingly accurately for arbitrary inelastic behavior as further information is added to the representation. A rigorous analysis of this question is beyond the scope of this paper and, instead, we have presented selected numerical examples that demonstrate the range of Data-Driven inelasticity and the numerical performance of implementations thereof.

A number of additional considerations and possible extensions of Data-Driven inelasticity suggest themselves.

\underline{\sl Connection to machine learning}.
We note that the closest-point projection $P_D$ in (\ref{K1TK7z}) entails a search over the entire material data set $D$. This search can be carried out, e.~g., by recourse to range-search algorithms \cite{Kirchdoerfer:2016}, tree-search algorithms, or similar fast search algorithms. Interestingly, search algorithms rely on spatial data structures, such as quadtrees and octrees, based on the principle of recursive subdivision. Such structures represent density, neighbor, clustering and other relations between the data points, which in turn may be regarded as a form of unsupervised machine learning (cf., e.~g., \cite{Duda:2001}). However, we emphasize that here the aim is to 'learn' the data set in its entirety, instead of replacing it by a model or some other reduced representation. In particular, the learning process does not entail any loss of information relative to the material data set.

\underline{\sl Multi-fidelity Data-Driven problems}.
A number of interrelated extensions and variations of the Data-Driven paradigm presented in this paper are noteworthy. We begin by noting that data enter the distance-minimizing Data-Driven problem (\ref{K5b6z0}) with uniform confidence, i.~e., all data are presumed to be equally reliable. However, in practice some data are of higher quality than others. The importance of keeping careful record of the pedigree, or ancestry, of each data point and of devising metrics for quantifying the level of confidence that can be placed on the data is well-recognized in Data Science \cite{Newman:2002, Raissi:2017}. A generalization of the distance-minimizing Data-Driven problem (\ref{yKEW6b}) that accounts for data fidelity is
\begin{equation}\label{97HKA5}
    \min_{\bz_{k+1} \in E_{k+1}}
    \min_{\by_{k+1} \in D_{k+1}}
    \Big(
        d^2(\bz_{k+1},\by_{k+1})
        +
        C(\by_{k+1}) ,
    \Big)
\end{equation}
where the fidelity cost $C(\by_{k+1}) \geq 0$ measures the uncertainty, or lack or fidelity, of data point $\by_{k+1}$. Thus, $C(\by_{k+1})=0$ if $\by_{k+1}$ is absolutely certain and $C(\by'_{k+1})\geq C(\by_{k+1})$ if $\by'_{k+1}$ is less certain, or of lesser fidelity, than $\by_{k+1}$. It is clear from (\ref{97HKA5}) that data points now influence the Data-Driven solution according to their fidelity, i.~e., high-fidelity data are given more weight in determining the solution than low-fidelity data.

A standard quantification of experimental data uncertainty consists of appending error bars to the data, corresponding to an estimate of the standard deviation of the measurements, and identifying the data points with the center of the distribution. If $s(\by_e)$ is the standard deviation of a local data point of mean value $\by_e$ and assuming a Gaussian distribution, the expected distance between a local state $\bz_e$ and the measurement is
\begin{equation}
    \int
        \frac{|\bz_e - \by'_e|^2}{(\sqrt{2\pi} s(\by))^{2m_e}}
        \exp\Big( - \frac{|\by_e-\by'_e|_e^2}{2 s^2(\by_e)} \Big)
    \, dy'_e
    =
    |\bz_e-\by_e|_e^2
    +
    2 m_e s^2(\by_e) .
\end{equation}
Comparing this identity with (\ref{97HKA5}) affords the identification
\begin{equation}
    C(\by) = \sum_{e=1}^M 2 m_e s^2(\by_e) ,
\end{equation}
which relates the fidelity cost $C(\by)$ to the uncertainty of the data.

\underline{\sl History-matching Data-Driven problems}.
The extended Data-Driven problem (\ref{97HKA5}) suggests the following variation of the Data-Driven inelasticity paradigm. Suppose that it is possible to collect history data of material elements directly, i.~e., a local material history repository $H_e$ is available consisting of corresponding pairs of short histories $\{\bz_{e,k+1-l}\}_{l=0}^{N}$ $=$ $(\{\mbs{\epsilon}_{e,k+1-l}\}_{l=0}^{N}, \{\mbs{\sigma}_{e,k+1-l}\}_{l=0}^{N})$. For instance, for the Standard Linear Solid, data repositories of this type consist of two-time histories $(\bz_{e,k}, \bz_{e,k+1})$ $=$ $(\{\mbs{\epsilon}_{e,k}, \mbs{\epsilon}_{e,k+1}\}, \{\mbs{\sigma}_{e,k}, \mbs{\sigma}_{e,k+1}\})$ in the local material history space $H_e = \mathbb{R}^{4 m_e}$. We can metrize $H_e$ by means of the norm
\begin{equation}
    | \{\bz_{e,k+1-l}\}_{l=0}^{N} |_e
    =
    \Big(
        \sum_{l=0}^{N} C_{e,l} | \bz_{e,k+1-l} |_e^2
    \Big)^{1/2} ,
\end{equation}
where $\{C_{e,l}\}_{l=0}^{N}$ are positive weights. We can further define a global material history set as $H = H_1 \times \cdots \times H_M$, with norm
\begin{equation}
    | \{\bz_{k+1-l}\}_{l=0}^{N} |
    =
    \Big(
        \sum_{e=1}^m w_e \, | \{\bz_{e,k+1-l}\}_{l=0}^{N} |_e^2
    \Big)^{1/2} .
\end{equation}
A history-matching Data-Driven problem can now be defined as
\begin{equation}\label{dcWC0i}
    \min_{\bz_{k+1} \in E_{k+1}}
    \min_{\{\by_{k+1-l}\}_{l=0}^{N} \in H}
        d^2(\{\bz_{k+1-l}\}_{l=0}^{N}, \{\by_{k+1-l}\}_{l=0}^{N}) ,
\end{equation}
i.~e., the Data-Driven solution at time $t_{k+1}$ is the admissible state $\bz_{k+1} \in E_{k+1}$ such that the history $\{\bz_{k+1-l}\}_{l=0}^{N}$ is closest to the material history set $H$. Thus, in this history-matching paradigm the prior history to $\by_{k+1}$ is no longer fixed to $\{\bz_{k+1-l}\}_{l=0}^{N-1}$ and all prior histories in $H$ are considered with weights depending on their distance to $\{\bz_{k+1-l}\}_{l=0}^{N-1}$. We further note that problem (\ref{dcWC0i}) is in fact of the form (\ref{97HKA5}) with cost
\begin{equation}
    C(\by_{k+1})
    =
    d^2(\{\bz_{k+1-l}\}_{l=0}^{N-1}, \{\by_{k+1-l}\}_{l=0}^{N-1}) .
\end{equation}
Thus, the history-matching reformulation of the Data-Driven problem simply collects into a data set $D_{k+1}$ all states $\by_{k+1}$ in $H$ and assigns them confidence weights according to the distance of the corresponding prior histories to the actual prior history $\{\bz_{k+1-l}\}_{l=0}^{N-1}$.

We have repeated the Standard Linear Solid test calculations described in Section~\ref{BuKeb1} using history material data sets in $((\epsilon_{e,k}, \sigma_{e,k}), (\epsilon_{e,k+1}, \sigma_{e,k+1}))$ space and history matching. The results of the calculations are ostensibly identical to those of Section~\ref{BuKeb1} and are not plotted here in the interest of brevity. History repositories enjoy the advantage that prior histories can be sampled off-line and a data set $D_{k+1}$ need not be known for all possible prior histories. The disadvantage is that history data add to the dimensionality of the data set. Therefore, history matching is only practical when prior histories are short, e.~g., in the context of low-order differential representations.

\underline{\sl Goal-oriented self-consistent data acquisition}.
An issue of critical importance concerns the acquisition of material data sets with appropriate coverage of phase space for specific applications. For general materials, phase space is of a dimension such that it cannot be covered uniformly by data. High-dimensional spaces are encountered in other areas of physics such as statistical mechanics, where the high dimensionality of state space is usually handled by means of {\sl importance sampling} techniques. The main idea is to generate data that are highly relevant to the particular problem under consideration, while eschewing irrelevant areas of phase space. A method for generating such goal-oriented data sets is the self-consistent approach of Leygue {\sl et al.} \cite{Leygue:2018}. In that approach, from a collection of non-homogeneous strain fields, e.~g., measured through Digital Image Correlation (DIC), a self-consistent iteration builds a material data set of strain--stress pairs that cover the region of phase-space relevant to a particular problem. In effect, the self-consistent approach generates the material data set and solves for the corresponding Data-Driven solution simultaneously.

These extensions and generalizations of Data-Driven inelasticity suggest worthwhile directions for further research.

\section*{Acknowledgments}

MO gratefully acknowledges the support of the Deutsche Forschungsgemeinschaft (DFG) through the Sonderforschungsbereich 1060 {\sl ``The mathematics of emergent effects''}. SR and RE gratefully acknowledge the financial support of the Deutsche Forschungsgemeinschaft (DFG) through the project {\sl ``Model order reduction in space and parameter dimension – towards damage-based modeling of polymorphic uncertainty in the context of robustness and reliability''} within the priority programm SPP 1886 {\sl ``Polymorphic uncertainty modelling for the numerical design of structures''}.

\bibliography{Biblio_clean}
\bibliographystyle{unsrt}

\end{document}